\documentclass[journal,twoside,web]{ieeecolor}
\usepackage{generic}
\usepackage{cite}
\usepackage{amsmath,amssymb,amsfonts}
\usepackage{graphicx}
\usepackage{textcomp}

\usepackage{bm}
\usepackage{mathrsfs}
\usepackage{color}
\usepackage{url}
\usepackage{longtable}
\usepackage{multirow}
\usepackage{algorithm}
\usepackage{algorithmicx}
\usepackage{algpseudocode}
\algrenewcommand{\algorithmiccomment}[1]{\hskip3em\% #1}

\usepackage{dblfloatfix} 

\usepackage[hang,small,bf]{caption}
\usepackage[subrefformat=parens]{subcaption}
\newtheorem{assumption}{Assumption}
\newtheorem{lemma}{Lemma}
\newtheorem{theorem}{Theorem}
\newtheorem{corollary}{Corollary}
\newtheorem{definition}{Definition}
\newtheorem{proposition}{Proposition}

\newtheorem{remark}{Remark}
\newtheorem{example}{Example}

\newcommand{\A}{\mathcal{A}}

\newcommand{\C}{\mathbb{C}}

\newcommand{\sC}{\mathscr{C}}

\newcommand{\E}{\mathcal{E}}

\newcommand{\I}{\mathcal{I}}

\newcommand{\cL}{\mathcal{L}}
\newcommand{\N}{\mathcal{N}}
\newcommand{\bN}{\mathbb{N}}
\newcommand{\sN}{\mathscr{N}}

\newcommand{\R}{\mathbb{R}}

\newcommand{\cS}{\mathcal{S}}

\newcommand{\sS}{\mathscr{S}}

\newcommand{\V}{\mathcal{V}}

\newcommand{\im}{\mathrm{im\hspace{0.2ex}}}
\newcommand{\rank}{\mathrm{rank\hspace{0.2ex}}}
\newcommand{\supp}[1]{{\rm supp}\left({#1}\right)}
\newcommand{\blocksupp}[1]{{\rm blocksupp}\left({#1}\right)}

\def\BibTeX{{\rm B\kern-.05em{\sc i\kern-.025em b}\kern-.08em
		T\kern-.1667em\lower.7ex\hbox{E}\kern-.125emX}}
\begin{document}
	\title{Distributed Resilient State Estimation and Control with Strategically Implemented Security Measures}
	
	\author{Takumi Shinohara, \IEEEmembership{Member, IEEE}, Karl Henrik Johansson, \IEEEmembership{Fellow, IEEE}, \\ and Henrik Sandberg, \IEEEmembership{Fellow, IEEE}
		\thanks{This work was supported in part by the Knut and Alice Wallenberg Foundation Wallenberg Scholar Grant and Swedish Research Council (Project 2023-04770).}
		\thanks{The authors are with the Department of Decision and Control Systems, KTH Royal Institute of Technology, and also with Digital Futures, 100 44 Stockholm, Sweden. (e-mail: \{tashin, kallej, hsan\}@kth.se).}}
	
	\maketitle
	
	\begin{abstract}
		This paper addresses the problem of distributed resilient state estimation and control for linear time-invariant systems in the presence of malicious false data injection sensor attacks and bounded noise. 
		We consider a system operator (defender) capable of deploying cybersecurity measures to counteract malicious sensor compromises.  
		Although such measures enhance resilience against adversarial attacks, they may incur substantial costs; hence, it is crucial to strategically select countermeasures that balance resilience gains and cost efficiency. 
		We first demonstrate that the system's resilience against attacks is maximized through the appropriate implementation of security measures, implying that no attacker can execute undetectable sensor attacks.
		Building on this analysis, we formulate an optimization problem for optimal security measures under cost constraints, which is NP-hard.
		We then propose an exact algorithm and derive a polynomial-time approximation algorithm that achieves a constant-factor performance guarantee.
		Furthermore, we develop a distributed resilient state estimation and control scheme informed by the optimal security measure and establish conditions that guarantee bounded estimation and control errors. 
		Finally, we validate the efficacy of our approach through numerical simulations of a vehicle platooning scenario.
	\end{abstract}
	
	\begin{IEEEkeywords}
		Resilient control systems, distributed state estimation, distributed control, sensor attacks.
	\end{IEEEkeywords}
	
	\section{Introduction}
	\label{sec:introduction}
	\IEEEPARstart{N}{etwork-based} distributed control systems have a wide range of applications and support various infrastructure systems, including power systems, intelligent transportation systems, and water supply systems. 
	However, as the scale of the systems increases and interactions between systems become more complex, they become attractive targets for malicious cyberattacks. 
	Such attacks pose a considerable threat to the reliability and resilience of these systems \cite{2013TACBullo,2015Johansson,2022Sandberg}.
	Indeed, there have been several cases of hostile cyberattacks against such systems that have caused economic loss and severe damage to facilities and human activities \cite{2024IEEEAccess,2021IEEEAccess}. 
	Consequently, establishing resilient estimation and control methods against malicious attacks in distributed control systems is imperative.
	
	In order to securely control distributed control systems in an adversarial environment, monitoring and estimating system states are essential.
	The distributed state estimation problem in the presence of attacks is often called \textit{distributed resilient state estimation} or \textit{distributed secure state estimation}, and it has been extensively studied in the past decade.
	For example, the work in \cite{2022TACJohanssson} addressed a linear system with bounded noise and proposed an algorithm based on a distributed saturation-based filter.
	The article \cite{2024TACMo} proposed a novel distributed secure state estimation scheme on a linear system subject to bounded noise, using decomposition and distributed convex optimization techniques.
	These papers have focused exclusively on the development of state estimation methods, while some other papers have proposed control methods in the presence of attacks.
	For instance, in \cite{2021AutomaticaJohansson}, the authors proposed a novel architecture comprising a resilient observer and an observer-based distributed controller for a group of vehicles modeled as homogeneous second-order dynamics.
	This architecture enables the vehicles to achieve accurate state estimation and formation control even if measurements from a subset of vehicle sensors are compromised by a malicious attacker.
	
	\subsection{Motivation and Contribution}
	This work addresses a distributed resilient state estimation and control problem under malicious false data injection sensor attacks.
	We consider linear time-invariant heterogeneous multi-agent systems subject to both process and measurement noise. 
	Existing studies commonly adopt the following assumptions: all sensors may be compromised, the system does not employ any countermeasure, the attacker can only compromise a partial set of sensors, and the network possesses sufficient sensor redundancy to withstand such attacks.
	While these assumptions yield valuable theoretical insights, real-world critical systems often incorporate specific cybersecurity countermeasures, such as anti-jamming techniques and intrusion detection/prevention systems.
	Motivated by this practical context, we consider the distributed resilient estimation and control problem under the assumption that the system operator can deploy security measures for some sensors and that the attacker has full knowledge and access to attack resources.
	Although these defenses improve resilience against these intimidating adversaries, they also incur substantial security costs; hence, countermeasures must be selected strategically, considering their cost-effectiveness.
	The primary contributions of this work are as follows:
	\begin{itemize}
		\item We show that the system's resilience, characterized by a security index, can be maximized by effectively selecting security measures. This means that no attacker, regardless of their knowledge and capabilities, can perform undetectable attacks.
		\item We present an algorithm that strategically determines the security measures for a system, considering the cost-effectiveness of these measures. This problem is NP-hard and hard to solve in general, so we give an approximation algorithm for efficient computation.
		\item We develop a resilient state estimation and control framework for systems equipped with these optimally chosen security measures, and we derive sufficient conditions that guarantee estimation and control errors remain within a prescribed bound.
		\item  We validate our approach using a vehicle-platooning case study, demonstrating that our framework maintains resilience against a larger number of sensor attacks compared to conventional methods.
	\end{itemize}
	
	In short, our proposed framework provides more resilient distributed state estimation and control algorithms against malicious sensor attacks than the approaches in \cite{2024TACMo,2022TACJohanssson}. 
	Furthermore, compared to \cite{2021AutomaticaJohansson}, our method can be applied to a wider range of heterogeneous multi-agent systems.
	

	\subsection{Organization}
	We first formulate the system model, observation model considering security measures, attacker model, and communication model in Section~\ref{section:problem}.
	Then, in Section~\ref{section:resilience_enhancement}, we demonstrate that the system's resilience can be maximized by implementing security measures appropriately.
	In Section~\ref{section:strategic_implementation}, we propose a strategic security implementation algorithm.
	The optimization problem is NP-hard, and hence we provide an approximation algorithm for efficient computation.
	We then develop a distributed resilient estimation and control framework considering the security measure, and we conduct a performance analysis of the proposed framework in Section \ref{section:distributed_algorithm}.
	Section~\ref{section:simulation} shows simulation results using a vehicle platooning model to corroborate the effectiveness of the proposed framework.
	Section~\ref{section:conclusion} finally concludes this article.
	
	\subsubsection*{Notations}
	The symbols $ \R $, $ \R^+ $, $ \C$, and $ \bN_0 $ denote the set of real numbers, positive real numbers, complex numbers, and nonnegative integers, respectively.
	The notation $ |\I| $ denotes the cardinality of a set $ \I $.
	For a vector, the support of the vector is defined as $ \supp{x} $.
	Also, given a block vector $ y \triangleq [y_1^\top ,\ldots, y_N^\top ]^\top $, we denote by $ \mathrm{blocksupp}\left(y\right) \triangleq \{i: y_i \neq 0 \} \subseteq \{1,\ldots,N\}$ the block support of $ y $.
	We use the notations $ \| x \| $ and $ \| A \| $ to denote the $ \ell_2 $ norm of a vector $ x $ and the induced $ \ell_2 $ norm of a matrix, respectively.
	$ A \otimes B $ is the Kronecker product of matrices $ A $ and $ B $.
	Given a linear map $ A $, we use $ \ker A $ to denote the kernel of $ A $.
	The sets of eigenvalues and eigenvectors of $ A $ are, respectively, denoted by $ \sigma(A) $ and $ \mu (A) $.
	The spectral radius of $ A $, i.e., the largest absolute value of its eigenvalues, is denoted by $ \rho(A) $. 
	The identity matrix with dimension $ n \times n $ is denoted as $ I_n $.
	The zero matrix with dimension $ m \times n $ is defined as $ 0_{m,n} $, and $ 0_n $ is used when the dimension is $ n \times n $ for simplicity.
	We use $ \bm{1}_n $ to indicate an $ n $-dimensional vector whose entries are $ 1 $.

	For a vector $ x \in \R^n $ and an index set $ \I \subseteq \{1,\ldots,n\} $, we use $ x_\I \in \R^{|\I|} $ to denote the subvector obtained from $ x $ by removing all elements except those indexed by the set $ \I $.
	Similarly, for a matrix $ A\in \R^{m \times n} $ and an index set $ \Gamma \subseteq \{1,\ldots,m\} $, we use $ A_\Gamma \in \R^{|\Gamma|\times n} $ to denote the submatrix obtained from $ A $ by removing all rows except those indexed by $ \Gamma $.
	A similar expression is applied to the block matrix $ M \triangleq [M_1^\top ,\ldots, M_N^\top ]^\top $.
	Hence, $ M_\Gamma $ is the matrix obtained from the block matrix $ M $ by eliminating all blocks except those indexed by $ \Gamma \subseteq \{1,\ldots, N\}$.
	For example, if $ M = [M_1^\top ,\ldots, M_5^\top ]^\top $ and $ \Gamma = \{1,3,4\} $, then we obtain $ M_\Gamma = \left[M^\top_1,M^\top_3,M^\top_4\right]^\top. $
	
	\section{Problem Formulation}
	\label{section:problem}
	In this section, we introduce the system model, followed by descriptions of the observation model that includes security measures, the attacker model, and the communication model. The problem of interest addressed in this paper is then presented at the end.

	\subsection{System Model}
	This paper deals with a heterogeneous multi-agent system consisting of $ N $ agents.
	Heterogeneous multi-agent systems naturally arise as models of cyber-physical systems across various application domains \cite{Hetro-01,Hetero-02}.
	The dynamics of each agent are given by
	\begin{align}
		\label{eq:Ori_State}
		x_i(k+1) = A_ix_i(k) + B_iu_i(k) + w_i(k),
	\end{align}
	where $ x_i(k) \in \R^{n_i} $, $ u_i(k) \in \R^{m_i} $, and $ w_i(k) \in \R^{n_i} $ denote, respectively, the 
	unknown system state, control input, and bounded process noise with $ \| w_i(k)\| \leq \delta^w_i$ of the $ i $th agent at time $ k $.
	Denote the set of all agents by $ \V \triangleq \{1,\ldots,N\} $.
	The overall dynamics of all agents can be obtained as
	\begin{align}
		\label{eq:overall_dynamics}
		x(k+1) = A x(k) + B u(k) + w(k),
	\end{align}
	where 
	\begin{align*}
		A&~\triangleq \mathrm{blkdiag}(A_1,\ldots,A_N)\in\R^{n\times n}, \\
		B&~\triangleq \mathrm{blkdiag}(B_1,\ldots,B_N)\in \R^{n \times m}, \\
		x(k) & \triangleq [~x_1(k)^\top,\ldots,x_N(k)^\top~]^\top \in \R^{n}, \\
		u(k) & \triangleq [~u_1(k)^\top,\ldots,u_N(k)^\top~]^\top \in \R^{m}, \\
		w(k) & \triangleq [~w_1(k)^\top,\ldots,w_N(k)^\top~]^\top \in \R^{n}.
	\end{align*}
	Here, $ n \triangleq \sum_{i \in \V} n_i $ and $ m \triangleq \sum_{i \in \V} m_i $.
	Denote the upper bound of the global process noise by $ \delta_w $, i.e., $ \| w(k) \| \leq \delta_w $.
	
	
	\subsection{Observation Model with Security Measure}
	Each agent is equipped with some sensors to observe the system state.
	However, sensors are vulnerable and their measurements can be compromised by malicious attackers (e.g., GPS jamming or spoofing attacks).
	To counter the attacks, in this paper, we assume that a security measure (e.g., anti-jamming or encryption) can be implemented for each agent.
	Adopting a security measure for all agents enhances the security level against malicious sensor attacks, but the cost will be excessive.
	When the attacker's capability is limited, thorough security measures may be unnecessary; rather, a cost-effective measure is often suitable. 
	Hence, the security measures for the agents should be determined strategically.
	
	In this paper, we call an agent that implements a security measure a \textit{secure agent}; otherwise, we call it a \textit{normal agent}.
	Define $ \sC \triangleq \{ \sN, \sS \} $ as the choice set for each agent, where $ \sS $ indicates that the agent is secure and $ \sN $ means that the agent is normal.
	Let $ \phi_i \in \sC $ be the type for the $i $th agent.
	The types for all agents are defined as $ \phi \triangleq [~\phi_1,\ldots,\phi_N~]^\top \in \sC^N $.
	Denote the index set of agents with $\phi_i = \sN $ (resp. $\phi_i = \sS $) by $\N \subseteq \mathcal{V} $ (resp. $\mathcal{S} \subseteq \mathcal{V}$).
	Note that $ \mathcal{V}= \mathcal{N} \cup \mathcal{S}  $ and determining $ \phi $ is equivalent to determining $ \cS $.
	For a secure agent, its sensor measurement is protected against attacks; however, the implementation cost is higher than that of the normal one.
	For each agent $ i $, we define the costs of choosing $ \sN $ and $ \sS $ as $ c^\mathcal{N}_i\in \R^+ $ and $c^\mathcal{S}_i \in \R^+ $, respectively, where $c^\mathcal{N}_i< c^\mathcal{S}_i $.
	Thus, the total security cost is 
	\begin{align}
		\label{eq:cost}
		c(\phi) \triangleq \sum_{i \in \mathcal{N}}c^\mathcal{N}_i  + \sum_{i \in \mathcal{S}} c^\mathcal{S}_i.
	\end{align}
	We sometimes use a notation $ c(\cS) $, which denotes the total security cost based on the secure agent set $ \cS \subseteq \V $, i.e., $ c(\cS) \triangleq \sum_{i \in \mathcal{\V \backslash \cS}}c^\mathcal{N}_i  + \sum_{i \in \mathcal{S}} c^\mathcal{S}_i$.

	Based on $\phi$, the output of each agent is given by
	\begin{align}
		\label{eq:y_i}
		y_i(k) & =  h_i \left(x(k),a_i(k),\phi_i\right) \nonumber \\
		& \triangleq \left\lbrace \begin{array}{ll}
			C_ix(k) + v_i (k)+ a_i(k), &\hspace{-1mm} \mathrm{if}~\phi_i = \sN, \\
			C_ix(k) + v_i(k), & \hspace{-1mm} \mathrm{if}~\phi_i = \sS ,
		\end{array}\right. 
	\end{align}
	where $ v_i(k) \in \R^{{p_i}} $ represents the bounded measurement noise with $ \| v_i(k)\| \leq \delta^v_i $, $ a_i (k) \in \R^{p_i} $ models the false data injection attack against the measurement designed by malicious adversaries, $ C_i \in \R^{p_i \times n} $ is the measurement matrix, and $ y_i \in \R^{p_i} $ indicates the output from sensors of the $ i $th agent.
	One can observe that the output of the secure agent is protected against the attack.
	We define the total number of measurements as $ p \triangleq \sum_{i\in\V} p_i $. 
	
	\begin{figure*}[t]
		\begin{center}
			\includegraphics[width=0.8\linewidth]{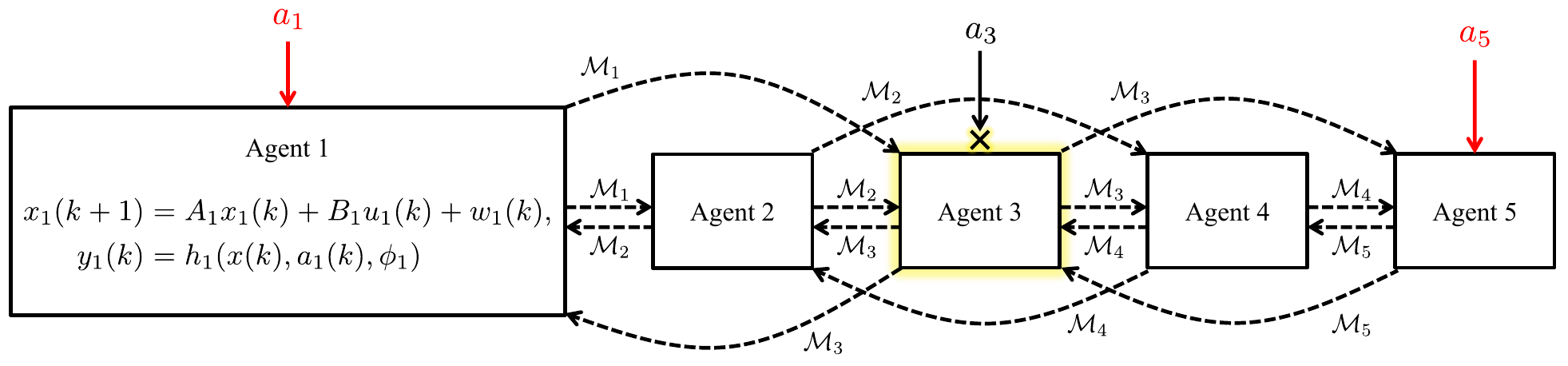}
			\caption{An example of a system architecture with five agents, where the third agent is the secure one.
				Agents 2 to 5 have dynamics similar to agent 1, based on $ A_i $ and $ B_i $.
				Each agent sends its message to its neighbors.
				A malicious attacker injects malicious signals to agents 1, 3, and 5, but the attack on the third agent is prevented by the security measure.}
			\label{fig:system_diagram_overall}
%
%
		\end{center}
	\vspace{-5mm}
	\end{figure*}
	
	All measurements are given by
	\begin{align}
		\label{eq:measurement_all}
		y(k) = h \left(x(k),a(k),\phi \right) & \triangleq \left[\begin{array}{c}
			h_1 \left(x(k),a_1(k),\phi_1\right) \\ \vdots \\ h_N \left(x(k),a_N(k),\phi_N\right)
		\end{array}\right] \nonumber \\
		& = Cx(k) +v(k)+ V(\phi) a(k),
	\end{align}
	where
	\begin{align}
		\label{eq:V(phi)}
		C & \triangleq [~C^\top_1,\ldots,C^\top_N~]^\top \in \R^{p \times n}, \nonumber\\
		y(k) & \triangleq [~y_1(k)^\top,\ldots,y_N(k)^\top~]^\top \in \R^{p}, \nonumber\\
		v(k) & \triangleq [~v_1(k)^\top,\ldots,v_N(k)^\top~]^\top \in \R^{p}, \nonumber\\
		a(k) & \triangleq [~a_1(k)^\top,\ldots,a_N(k)^\top~]^\top \in \R^{p}, \nonumber\\
		V(\phi) & \triangleq \mathrm{blkdiag}\left(V_1(\phi_1),\ldots,V_N(\phi_N)\right) \in \R^{p\times p}, \nonumber\\
		V_i(\phi_i)  & \triangleq \left\lbrace \begin{array}{ll}
			I_{p_i}, & \mathrm{if}~\phi_i = \sN, \\
			0_{p_i}, & \mathrm{if}~\phi_i= \sS.
		\end{array}\right.
	\end{align}
	Denote the upper bound of the global measurement noise by $ \delta_v $, i.e., $ \| v(k) \| \leq \delta_v $.
	We assume that $(A,C)$ is observable, but we do not assume that each pair $ (A,C_i) $ is observable for any $ i \in \V $.
	\begin{example}
		Consider $ N = 3 $ and $ p_1 = p_2 = p_3 = 1 $ (i.e., $ p = 3 $).
		If only the first and third agents implement a security measure, namely, $ \phi = \left[\sS, \sN, \sS\right]^\top $, then all measurements are formulated as
		\begin{align*}
			y(k) = Cx(k) + v(k) + \left[\begin{array}{ccc}
				0 & & \\ & 1 &  \\ & & 0
			\end{array}\right]a(k),
		\end{align*} 
		which implies the sensor attack against the first and third agents is protected by the security measure.
		In that case, the total security cost is given by $ c(\phi) = c_1^\cS + c_2^\N + c_3^\cS $.
	\end{example}
	
	\subsection{Attacker Model}
	A malicious attacker is assumed to be omniscient, namely, he/she possesses complete knowledge of the system dynamics, states, inputs, and outputs. 
	Also, it is assumed that the security measure $ \phi $ is known to the attacker.
	Based on the knowledge, the adversary is able to generate an attack sequence $ a(k) $ arbitrarily in terms of stochastic properties, bounds of magnitude, and time correlations like covert attacks or replay attacks.
	For an attack sequence $a=\{a(k)\}_{k \in \mathbb{N}_0}$ with security measure $\phi$, let $\mathrm{blocksupp}_{\phi}(a) \triangleq \bigcup_{k \in \mathbb{N}_0} \blocksupp{V(\phi)a(k)} $.
	
	\begin{remark}
		\label{remark:security_measure}
		Generally speaking, the security measure $ \phi $ is confidential information and should be protected from being disclosed to malicious adversaries. 
		Also, the assumption that the attacker knows the system model will be to his/her advantage. 
		In this paper, we assume that the attacker has knowledge of these, which is a tough assumption for the defense side. 
		Nevertheless, it should be noted that if the system deploys the security measure appropriately, there is no chance for the attacker to design stealth attacks even under this assumption, as confirmed in the next section.
	\end{remark}
	
	\begin{remark}
		\label{eq:number_attacks}
		Many previous studies have made assumptions regarding the number of attacks.
		Specifically, when the maximum number of compromised outputs is defined as $ l $, it is assumed that $ l $ is less than half the number of all outputs.
		This assumption arises from the need to ensure sensor redundancy in order to achieve secure state estimation.
		In contrast, we do not impose such an assumption.
		This is because, by appropriately adopting the security measure for the system, even if an attacker can carry out malicious sensor attacks exceeding the above thresholds, the attack can be detected (i.e., there are no undetectable attacks).
		This result can also be found in the next section.
	\end{remark}

	\subsection{Communication Model}
	We model the agent communication topology by an undirected connected graph $ \mathcal{G} = (\mathcal{V}, \mathcal{E}) $, where $ \V $ and $ \E \subseteq \V\times \V $ are the set of agents and edges, respectively.
	If there is an edge $ (i,j)\in \E $, the $ i $th agent can exchange information with the $ j $th agent.
	The Laplacian matrix of the graph is denoted by $ \mathcal{L} $.
	The $ i $th eigenvalue of $ \cL $ is defined as $ \lambda_i^\cL $ with $ \lambda_1^\cL\leq \lambda_{2}^\cL \leq \cdots\le\ \lambda_{\max}^\cL $.
	In an undirected connected graph, it is well-known that $ \lambda_1^\cL = 0 $ and $ \lambda_2^\cL > 0 $.
	We denote the neighborhood set of the $ i $th agent by $ \N_i  \triangleq \{j \in \V:(i,j) \in \E \}$.
	Each agent $ i \in \V $ is able to send a message to its neighboring agent $ j \in \N_i $ at time $ k $, denoted by $ \mathcal{M}_i(k) $.
	We assume that all agents are non-Byzantine, i.e., they generate the messages prescribed by the distributed algorithm.
	In addition, the inter-agent communication channels are assumed to be authenticated and integrity-protected, so that the message received by a neighbor coincides with the transmitted one.
	Accordingly, the attack surface considered in this paper is restricted to sensor channels of normal agents.
	The specific message content will be provided later.
	
	An example of system architecture we are interested in is depicted in Fig.~\ref{fig:system_diagram_overall}. 
	
	\subsection{Problem of Interest}
	\label{subsection:problems}
	In this paper, we aim to tackle the following problems.
	\begin{enumerate}
		\item \textit{Design a distributed resilient state estimator and controller executed in each agent by employing a given security measure $ \phi_i $, potentially compromised sensor output $ y_i(k) $, and the received neighborhood messages over the communication network $ \mathcal{G} $ such that}
		\begin{align}
			\label{eq:evaluation_function}
			\limsup_{k \rightarrow \infty}\! \dfrac{1}{N} \!\sum_{i\in\V} \!\left(\left\| \hat{x}_i(k) \!-\! x_i(k)\right\|  \!+\! \left\| x_i(k) \!-\! x^*_i(k)\right\|\right)\! \leq \!\Delta,
		\end{align}
		\textit{where $ \hat{x}_i(k) \in \R^{n_i} $ is the state estimate, $ x^*_i(k) \in \R^{n_i}$ is the desired state of the $ i $th agent at time $ k $, and $ \Delta \geq 0 $ is a scalar reflecting the performance of the proposed algorithm.}
		\item \textit{To achieve this, first determine the optimal security measure $ \phi^\star $ strategically to maximize the system resilience against malicious sensor attacks within a budget constraint.}
	\end{enumerate}
	
	Note that the security measures for all agents are centrally determined by the system operator. Based on the optimal measures, a resilient state estimator and controller is designed, which is executed in a distributed manner by each agent.
	
	\section{Security Index and Measure}
	\label{section:resilience_enhancement}
	Before tackling the above problems, we first show how the resilience of the overall system can be maximized by implementing $ \phi $ appropriately.
	To present this, we adopt a \textit{security index} of the system as a resilience metric.
	The undetectability of an attack defines this index, so we first provide a formal definition of an undetectable attack based on \cite[Lemma 3.1]{2013TACBullo}.
	\begin{definition}
		\label{def:undetactable_attack}
		The nonzero sensor attack sequence $ a=\{a(k)\}_{k \in \mathbb{N}_0} $ is \textit{undetectable} if and only if $ y(x^1(0), u, a, \phi, k) = y(x^2(0), u, 0, \phi, k),~\forall k \in \mathbb{N}_0 $ for distinct initial states $ x^1(0) $ and $ x^2(0) $, where $ y(x,u,a,\phi,k) $ denotes the sensor outputs at time $ k $ of the entire system with security measure $ \phi $ generated from the initial state $ x $, control input $ u $, and the attack injection $ a $.	
	\end{definition}
	
	In other words, an attack is undetectable if the sensor outputs generated by the attack coincide with the outputs of some nominal operating condition.
	Given this notion of undetectable attacks, the security index of the system considering the security measure is defined as follows \cite{2016CDCChong,2019AutomaticaChong}.
	\begin{definition}
		\label{def:security_index}
		The \textit{security index} of the overall system with security measure $ \phi $, denoted by $ \alpha_\phi(A, C) $, is defined as the minimum number of outputs (agents) that need to be compromised by an attacker to conduct an undetectable attack.
	\end{definition}
	
	The security index $ \alpha_\phi(A, C) $ quantifies resilience against stealthy sensor-channel attacks.
	Note that it does not capture falsification of inter-agent communication messages.
	According to the definition, the security index can be obtained by solving the following optimization problem:
	\begin{align}
		\label{eq:security_index}
		\alpha_\phi(A,C) &= \min_{a \neq 0} \left| \mathrm{blocksupp}_{\phi}(a) \right|, \\
		\label{eq:security_index_constraint}
		\mathrm{s.t.}&~~ y(x^1(0), u, a, \phi, k) = y(x^2(0), u, 0, \phi, k), \nonumber \\
		&~~\forall k \in \mathbb{N}_0,~\mathrm{for~some~distinct~}x^1(0) ,~x^2(0) \in \R^{n}.
	\end{align}
	One can observe that a smaller security index is more advantageous for an attacker since it means the attacker requires fewer resources to carry out an undetectable attack.
	On the other hand, enhancing this index increases the number of sensors that must be compromised to achieve undetectable attacks, which is advantageous for the system operator.
	The higher this index, the higher the system's resilience against undetectable attacks; hence, this index characterizes the system's resilience.
	Note that the above optimization problem may not have a solution; that is, there may be cases where adversaries cannot design any undetectable attacks.
	In that case, we adopt $ \alpha_\phi(A, C) = +\infty $.
	The possible range of this index is given as $ \alpha_\phi(A, C) \in \{1,\ldots,N\}\cup \{+\infty\}. $
	Note that the computation of the security index is NP-hard in general \cite{2014TACSandberg,2021AutomaticaSandberg}.
	
	The following theorem provides a necessary and sufficient condition so that attackers cannot design any undetectable attacks, namely, $ \alpha_\phi(A,C) = +\infty $, in the presence of an appropriate security measure $ \phi $.
	\begin{theorem}
		\label{theorem:security_index}
		For the system with security measure $ \phi $, an attacker cannot design any undetectable attacks (i.e., $  \alpha_\phi(A,C) = +\infty  $) if and only if $ (A, C_\cS) $ is observable, where $ C_\cS $ denotes the submatrix obtained from $ C $ by removing all blocks except those indexed by $ \cS $.
	\end{theorem}
	\begin{proof}
		See Appendix \ref{appendix:proof_theorem1}.
	\end{proof}
	
	This theorem shows that if the system operator appropriately implements the security measures such that $ (A, C_\cS) $ is observable, then the security index is maximized and there is no chance for an attacker to design undetectable attacks.
	It should be noted that this condition does not depend on the attack property.
	More specifically, even if the adversary is capable of compromising all sensor outputs, the system operator can prevent any undetectable attacks by deploying the security measure such that $ (A, C_\cS) $ is observable.
	On the other hand, if the observability of $ (A, C_\cS) $ cannot be achieved due to a budget constraint, then there exists an undetectable attack for the system.
	Therefore, strategically determining $ \phi $ under a budget constraint is essential, and the corresponding procedure will be presented in the next section.
	
	For future analysis, we provide an alternative characterization of the existence of undetectable attacks.
	For each eigenvalue $ \lambda $ of $ A $, define the global eigenspace:
	\begin{align*}
		\mathcal{E}(\lambda) \triangleq \ker \left(\lambda I - A\right) = \bigoplus_{i \in \V} \Pi_i \underbrace{\ker(\lambda I - A_i)}_{\mathcal{E}_i(\lambda)},
	\end{align*}
	where $ \oplus $ denotes the direct sum of subspaces in $ \C^n $, $ \E_i(\lambda) $ denotes the eigenspace of $ A_i $, and $ \Pi_i \in \R^{n\times n_i} $ denotes a block-selection matrix that embeds a local vector $ z \in \C^{n_i} $ into the global state space $ \C^n $ by placing it in the $ i $th block and zeros elsewhere, that is,
	\begin{align*}
	\Pi_ i \triangleq \left[\begin{array}{ccccccc}
			\!\!0_{n_1, n_i}^\top \!& \!\cdots\! & \!0_{n_{i-1}, n_i}^\top\! &\! I_{n_i}\! &\! 0_{n_{i+1}, n_i}^\top \!& \!\cdots\! & \!0_{n_N, n_i}^\top\!\!
		\end{array}\right]^\top\!\!.
\end{align*}
	The set of nonzero eigenvectors of $A$ is denoted by $\mu(A) \triangleq \bigcup_{\lambda\in\sigma(A)} (\mathcal{E}(\lambda)\setminus\{0\})$.
	Then, for the existence of undetectable attacks, we have the following proposition.
	\begin{proposition}
		\label{proposition:undetectable_attacks}
		For the system with security measure $ \phi $, the following statements are equivalent:
		\begin{enumerate}
			\item An undetectable attack exists.
			\item $ \ker C_\cS \cap \E(\lambda) \neq \{0\} $ for some $ \lambda \in \sigma(A) $.
			\item The security index is finite and equals $ \alpha_\phi (A,C) = \min_{v \in \mu(A)}\left| \left\lbrace i \in \mathcal{N} : C_{\{i\}\cup \cS } v \neq 0 \right\rbrace \right|$. 
		\end{enumerate}
	\end{proposition}
	\begin{proof}
		See Appendix~\ref{appendix:proof_proposition:undetectable_attacks}.
	\end{proof}


	Assuming a resource-limited adversary, this proposition establishes the following result.
	\begin{corollary}
		\label{corollary:number_attacks}
		Suppose that an attacker can compromise at most $ l $ agents, 
		i.e., any admissible attack sequence \(a\) satisfies $ |\mathrm{blocksupp}_{\phi}(a)| \leq l$.
		For the system with security measure $ \phi $, an undetectable attack exists if and only if $ l \geq \alpha_{\phi}(A,C) $.
	\end{corollary}
	\begin{proof}
		See Appendix \ref{appendix:proof_corollary:number_attacks}.
	\end{proof}

This corollary implies that an attacker cannot execute an undetectable attack if they can compromise only strictly fewer than $ \alpha_{\phi}(A,C) $ outputs.
Therefore, even when the condition $ \alpha_{\phi}(A,C) = + \infty $ cannot be achieved due to a budget constraint, the attack resilience of the system can be reinforced by appropriately employing security measures.
The system is protected against any attack compromising strictly fewer than $\alpha_\phi(A,C)$ outputs; equivalently, against at most $\alpha_\phi(A,C)-1$ outputs when $\alpha_\phi(A,C)<+\infty$.
	
	\section{Strategic Implementation of $ \phi $}
	\label{section:strategic_implementation}
	This section addresses the second problem presented in Subsection \ref{subsection:problems}, that is, we aim here to determine the optimal security measure $ \phi $ to maximize the system resilience against malicious sensor attacks within a budget constraint.
	We first formulate an optimization problem to determine the optimal security measure for all agents, and then present exact and polynomial-time approximation algorithms to solve it.
	
	\subsection{Optimization Problem}
	Using the security index as a metric of the system's resilience, the problem for the optimal security measure can be written as the following optimization problem:
	\begin{align}
		\label{eq:implementation}
		P_0: \max_{\phi \in \mathscr{C}^N}~\alpha_\phi(A,C) ~~	\mathrm{s.t.}~c(\phi) \leq \beta,
	\end{align}
	where $ \beta \in \R^+ $ denotes the budget.
	One can interpret this problem as determining $ \phi $ that maximizes the security index within the budget constraint.
	Borrowing the result from the previous section, one can confirm that this problem is solved by any security measure $ \phi $ such that $ (A,C_\cS) $ is observable and $ c(\phi) \leq \beta $.
	As with the calculation of the security index, this optimization problem is generally NP-hard and cannot be solved in polynomial time unless $\mathrm{P} = \mathrm{NP}$.
	The algorithm we present in the next subsection, therefore, is computationally intensive.
	However, in Subsection \ref{subsection:efficient}, we also present a computationally efficient approximation algorithm.
	
	\subsection{Brute-force Search Algorithm}

	\begin{algorithm}[t]
		\caption{Brute-force search algorithm for optimal security measure $ \phi^\star $}
		\label{algorithm:centralized}
		\begin{algorithmic}[1]
			\Require $ A,C, c(\cdot), \beta $, $ V(\cdot) $
			\Ensure $ \phi^\star $
			\Statex \hspace{-5.8mm}[Step 1] \Comment{Find $ \phi $ such that $ (A,C_\cS) $ is observable and $ c(\phi) \leq \beta $}
			\State $ \cS^\star \leftarrow \emptyset $
			\State $ c^\star  \leftarrow \infty $
			\For{$ l=1 $ to $ N $}
			\For{all $ \cS  \subseteq \V $ with $ |\cS| = l $}
			\If{$ c(\phi) \leq \beta $}
			\State $ \bar c \leftarrow c(\phi) $	\Comment{provisional cost}
			\If{$ (A,C_\cS) $ is observable and $ \bar c < c^\star $}
			\State $ \cS^\star \leftarrow \cS $ and  $ c^\star \leftarrow \bar c $
			\EndIf
			\EndIf
			\EndFor
			\EndFor
			\If{$ \cS^\star = \emptyset $}
			\Statex \hspace{-5.8mm}[Step 2] \Comment{Find $ \phi $ that makes the security index as large as possible subject to $ c(\phi) \leq \beta $}
			\State $ \alpha^\star \leftarrow 0 $
			\For{$ l=0 $ to $ N-1 $}
			\For{all $ \cS  \subset \V $ with $ |\cS| = l $}
			\If{$ c(\phi) \leq \beta $}
			\State $ \bar c \leftarrow c(\phi) $	\Comment{provisional cost}
			\State Calculate 
			\Statex \hspace{22mm}$ \alpha =  \min_{v \in \mu(A)}\left| \left\lbrace i \in \mathcal{N} : C_{\{i\}\cup \cS } v \neq 0 \right\rbrace \right| $
			\If{$\alpha > \alpha^*$ or ($ \alpha = \alpha^\star $ and $ \bar c < c^\star $)}
			\State	$ \cS^\star \leftarrow \cS $, $ \alpha^\star \leftarrow \alpha $, and $ c^\star \leftarrow \bar c $
			\EndIf
			\EndIf
			\EndFor
			\EndFor
			\EndIf
			\State \textbf{return} $ \phi^\star $ based on $ \cS^\star $
		\end{algorithmic}
	\end{algorithm}
	
	The brute-force search algorithm for the optimal security measure $ \phi^\star $ is given in Algorithm~\ref{algorithm:centralized}.
	This algorithm consists of two steps.
	In the first step, the algorithm seeks $ \phi $ such that $ (A,C_\cS) $ is observable and $ c(\phi) \leq \beta $.
	In other words, by calculating the cost and observability of $ (A, C_\cS) $ for all subsets $ \cS \subseteq \mathcal{V}$, it yields $ \phi $ that satisfies the budget constraint and for which no undetectable attacks exist.
	If the algorithm returns any solution in the first step, we can obtain the optimal security measure for the system that achieves $ \alpha_\phi(A,C) = +\infty $.
	If not, it is impossible to achieve $ \alpha_\phi(A,C) = +\infty $ within the budget constraint, and then we move to the second step.
	
	The second step aims to find $ \phi^\star $ within the budget constraint that maximizes the security index as much as possible.
	Since now there exists an undetectable attack, from Proposition~\ref{proposition:undetectable_attacks}, the problem of the security index (\ref{eq:security_index}) can be written as
	\begin{align}
		\label{eq:alpha_undetectable}
		\alpha_\phi (A,C) = \min_{v \in \mu(A)}\left| \left\lbrace i \in \mathcal{N} : C_{\{i\}\cup \cS } v \neq 0 \right\rbrace \right|.
	\end{align}
	As with the Step 1, by calculating the security index using this problem for all subsets $ \mathcal{S} \subset \V $ such that $ c(\phi) \leq \beta $, Step 2 of Algorithm~\ref{algorithm:centralized} provides $ \phi^\star $ under the situation where there exists an undetectable attack.
	Algorithm~\ref{algorithm:centralized} performs an exhaustive search over all secure agent sets and thus has worst-case complexity $ O(2^N) $ with respect to $ N $.

	\begin{remark}
		In this work, we employ the linear cost function $ c(\phi) $ and binary security decision for each agent (i.e., $ \phi_i $ is $ \sS $ or $ \sN $).
		Even if we consider a nonlinear cost function, we can incorporate it directly into the constraint of $ P_0 $. 
		In that case, the optimal security measure under a budget constraint can be obtained using Algorithm~\ref{algorithm:centralized} by enforcing the nonlinear budget constraint.
		The resulting measure can be utilized in the resilient estimation and control scheme discussed in Section~\ref{section:distributed_algorithm}.
		If multiple protection levels (e.g., basic, advanced, hardened) per agent are considered, after defining a corresponding security index for each configuration, a similar exhaustive search algorithm can be constructed.
		A systematic treatment of multi-level security models is left as future work.
	\end{remark}
	
	\subsection{Approximation Algorithm}
	\label{subsection:efficient}

	In this subsection, we present an approximation algorithm for $ P_0 $ that can be solved in polynomial time.
	We first provide the algorithm using a different characterization of the attack undetectability.
	We then show its approximation performance.
	
	\subsubsection{Preliminaries}
	For each agent $ i $, let $ U_i(\lambda) \in\C^{n_i \times r_i(\lambda)}$ denote a basis of the eigenspace for $ \lambda \in \sigma(A) $, namely, it holds that $ \im U_i(\lambda) = \E_i(\lambda) $.
	For each $ \lambda \in \sigma(A) $, define
	\begin{align*}
		 E_\lambda \!\triangleq \!\left[~\Pi_1 U_1(\lambda),\cdots,\Pi_N U_N(\lambda)~\right] \!\in \!\C^{n \times r_\lambda},~ r_\lambda \! \triangleq\! \sum_{i \in \V} r_i(\lambda).
	\end{align*}
	From the definition, it follows that $ \im E_\lambda = \E(\lambda) $.
	Since $ \Pi_i $ has full column rank, we have $ \rank \Pi_i U_i(\lambda)  = r_i(\lambda),~\forall i \in \V$, which implies $ \rank E_\lambda = \sum_{i \in \V} \rank \Pi_i U_i(\lambda)  = \sum_{i \in \V}r_i(\lambda) = r_\lambda$.
	This implies $ E_\lambda $ has full column rank as well.
	With these definitions, the condition for the existence of an undetectable attack can be rewritten as follows.
	\begin{lemma}
		\label{lemma:undetectable}
		For the system with security measure $ \phi $, $ \alpha_\phi(A,C) = +\infty $ if and only if $ \rank C_\cS E_\lambda = r_\lambda $ for all $ \lambda \in \sigma(A) $.
	\end{lemma}
	\begin{proof}
		See Appendix~\ref{appendix:proof_lemma:undetectable}.
	\end{proof}
	
	The condition $ \rank C_\cS E_\lambda = r_\lambda $ for all $ \lambda \in \sigma(A) $ means that all directions of each eigenmode are observable from the protected measurements.
	Hence, this lemma states that the security index is maximized if and only if every eigenmode is fully covered by the secure agents.
	In this case, an attacker, even if they can arbitrarily manipulate all normal outputs, cannot hide any behavior from the protected measurements.
	From this lemma, we can check whether the security index is maximized based on the rank condition of $ C_\cS E_\lambda $.

	For simplicity, we define $ R_\lambda(\cS) \triangleq \rank C_\cS E_\lambda $ for a given secure agent set $ \cS\subseteq\V $ and $ \lambda \in \sigma(A) $.
	Also, for $ \lambda \in \sigma(A) $ and $ \cS\subseteq \V $, define
	\begin{align}
		\label{eq:delta_lambda}
		\delta_\lambda(\cS) \triangleq r_\lambda - R_\lambda(\cS).
	\end{align}
	For each eigenvalue $ \lambda $, $ \delta_\lambda $ counts the number of unobservable directions.
	By Lemma~\ref{lemma:undetectable}, $ \alpha_\phi(A,C) = +\infty $ if and only if $ \delta_\lambda(\cS) = 0 $ for all $ \lambda \in \sigma(A) $.
	Unlike the security index $ \alpha_\phi(A,C) $, which may take the value $ + \infty $, the quantity $ \delta_\lambda(\cS)  $ is always finite and decomposes additively over the eigenvalues, which makes it suitable for quantifying how much performance is lost when we replace the optimal design by an approximate one.
	Note that $ r_\lambda $ does not depend on $ \cS $ (namely, a security measure $ \phi $), and hence maximizing
	\begin{align}
		\label{eq:R_S}
		R(\cS) \triangleq \sum_{\lambda \in \sigma(A)} R_\lambda(\cS)
	\end{align}
	directly corresponds to driving all $ \delta_\lambda(\cS) $ to zero.
	Since the matrix rank function $ R_\lambda(\cS) $ is known to be submodular and nondecreasing (see, e.g., \cite{Submodular}), $ R(\cS) $ is also submodular and nondecreasing.
	
	\subsubsection{Algorithm}
	
	\begin{algorithm}[t]
		\caption{Algorithm for approximate security measure $ \phi^a $}
		\label{algorithm:approximate}
			\begin{algorithmic}[1]
				\Require $ A, C, R_\lambda(\cdot), R(\cdot), \delta_\lambda(\cdot), E_\lambda,c(\cdot),c_i^\cS,c_i^\mathcal{N},\beta $
				\Ensure $\phi^a $
				\State $ \mathcal{A}_{12}\leftarrow\left\lbrace \cS \subseteq \V : |\cS| \leq 2,~c(\cS) \leq \beta \right\rbrace $
				\State $ \cS^a_{12} \leftarrow \arg \max_{\cS \in \A_{12}} R(\cS) $
				\State $ \mathcal{A}_3\leftarrow\left\lbrace \cS \subseteq \V : |\cS| = 3,~c(\cS)\leq \beta \right\rbrace $
				\For{each $ \cS \in \A_3 $}
				\State Initialization: $  \bar c \leftarrow c(\cS) $
				\State $ \Lambda(\cS) \leftarrow \{\lambda \in \sigma(A):\delta_\lambda(\cS)>0\} $
				\While{$ \cS \neq \V$, $  \bar c \leq \beta $, and $ \Lambda(\cS) \neq \emptyset $}
				\For{each $ i \in \V \backslash \cS $}
				\State Calculate \begin{align*}
					\hspace{10mm}\Delta_{(\cS,i)} =\!\! \sum_{\lambda \in \sigma(A)}\!\! \min \left\lbrace \delta_\lambda(\cS),\Delta R_{\lambda}(\cS\cup\{i\})\right\rbrace
				\end{align*}
				\EndFor
				\State $ i^* \leftarrow \arg \max_{i \in \V \backslash \cS} \dfrac{\Delta_{(\cS,i)}}{c_i^\cS - c_i^\mathcal{N}}$
				\State $ \cS \leftarrow \cS \cup \{i^*\} $
				\State $  \bar c \leftarrow c(\cS) $
				\State $ \Lambda(\cS) \leftarrow \{\lambda \in \sigma(A):\delta_\lambda(\cS)>0\} $
				\EndWhile
				\EndFor
				\State $ \cS^a_{3} \leftarrow \arg \max_{\cS \in \A_3} R(\cS) $
				\If{$ R(\cS_3^a) \geq R(\cS^a_{12}) $}
				\State $ \cS^a \leftarrow  \cS_3^a$
				\Else
				\State $ \cS^a \leftarrow  \cS_{12}^a$
				\EndIf
				\State \textbf{return} $ \phi^a $ based on $ \cS^a $
		\end{algorithmic}
	\end{algorithm}

	Consider Algorithm~\ref{algorithm:approximate}, which is based on an algorithm in \cite{Sviridenko} and returns an approximate security measure $  \phi^a $.
	The basic idea of this algorithm is as follows: Within the budget constraints, greedily implement a security measure to the most cost-effective agent such that $ \delta_\lambda(\cS^a) $ approaches zero, where $ \cS^a \subseteq \V $ denotes the approximate secure agent set.
	
	First, in Line 1, enumerate all feasible secure agent sets of cardinality one or two and, in Line 2, let $ \cS_{12}^a $ be a feasible set of cardinality one or two that has the largest value of the function $ R(\cdot) $.
	Then, in Line 3, let $ \A_{3} $ be the set of all feasible secure agents of cardinality three\footnote{The choice of three follows \cite{Sviridenko} and ensures the approximation guarantee established there. In practice, a smaller maximal cardinality can be used to further reduce the computational cost at the price of a weaker approximation guarantee.}.
	Subsequently, for each $ \cS \in \A_3 $, perform iterations after initialization, where $ \bar c $ denotes the provisional cost based on the iterated candidate set $ \cS $.
	Regarding Line 9, for $ \lambda \in \sigma(A) $ and $ \cS \subseteq \V $, we define
	\begin{align*}
		\Delta R_\lambda(\cS\!\cup\! \{i\}) & \triangleq R_\lambda(\cS\cup\{i\}) - R_\lambda(\cS), \\
		\Delta \delta_\lambda(\cS\!\cup\!\{i\}) & \triangleq \delta_\lambda(\cS) - \delta_\lambda(\cS\cup\{i\})\\
		& = \delta_\lambda(\cS) \!-\! \max\{0, \delta_\lambda(\cS) \!-\! \Delta R_\lambda(\cS\!\cup\! \{i\}) \} \\
		& = \min \{\delta_\lambda(\cS), \Delta R_\lambda(\cS\cup \{i\})\},
	\end{align*}
	namely, $ \Delta \delta_\lambda(\cS\cup\{i\}) $ denotes the reduction in (\ref{eq:delta_lambda}) resulting from implementing a security measure for agent $ i $.
	By summation over all $ \lambda \in \sigma(A) $, we have
	\begin{align*}
		\Delta_{(\cS,i)}  \triangleq \sum_{\lambda \in \sigma(A)}\!\! \min \left\lbrace \delta_\lambda(\cS),\Delta R_{\lambda}(\cS\cup\{i\})\right\rbrace
	\end{align*}
	which quantifies the effect of implementing a security measure for agent $ i $.
	Then, in Line 11, choose the most cost-effective agent $ i^* $ and, in Lines 12--14, update the candidate set $ \cS $, the provisional cost $ \bar c $, and the feasibility set $ \Lambda(\cS) $, respectively.
	Repeat these iterations until the candidate set $ \cS $ coincides with $ \V $, the budget is exceeded, or $ \Lambda(\cS) = \emptyset $, i.e., $ \delta_\lambda(\cS) = 0 $ for all $ \lambda \in \sigma(A) $.
	After the iteration, let $ \cS_{3}^a $ be a feasible set in $ \A_3 $ that has the largest value of the function $ R(\cdot) $.
	Finally, in Lines 18--22, the algorithm determines $ \cS^a $ as $ \cS^a_{3} $ if $  R(\cS_3^a) \geq R(\cS^a_{12})  $ and $ \cS^a_{12} $ otherwise, and in Line 23, return $ \phi^a $ based on $ \cS^a $.
%
	
	Algorithm~\ref{algorithm:approximate} is a knapsack-constrained greedy algorithm and requires $ O(N^5) $ function value computations \cite{Sviridenko}; thus, this algorithm runs in polynomial time.
	When Algorithm~\ref{algorithm:approximate} returns $ \phi^a $ satisfying $ c(\phi^a) \leq \beta $ and $ \delta_\lambda(\cS^a) = 0 $ for all $ \lambda \in \sigma(A) $, this $ \phi^a $ yields $ \alpha_{\phi^a}(A, C) = +\infty $ within the budget constraint, namely, $ \phi^a $ is an optimal solution of $ P_0 $.
	These conditions can be checked in polynomial time.
	Consequently, the computation of $ \phi^a $ via Algorithm~\ref{algorithm:approximate} and the verification of whether it is an optimal solution to $ P_0 $ can be carried out in polynomial time.

	\subsubsection{Approximation Performance}
	We next show the approximation performance of Algorithm~\ref{algorithm:approximate}.
	To present this, we consider the following metric:
	\begin{align}
		\label{eq:similarity}
		s(\phi^\star, \phi^a) \triangleq \sum_{\lambda \in \sigma(A)}\left(\delta_\lambda(\cS^a)-\delta_\lambda(\cS^\star)\right),
	\end{align}
	where $ \phi^\star $ is the solution of Algorithm~\ref{algorithm:centralized} and $ \cS^\star $ is the secure-agent set induced by $ \phi^\star $.
	Under the premise that $ \alpha_{\phi^\star}(A,C) = + \infty $, we have $ \delta_\lambda(\cS^\star) = 0 $ for all $ \lambda \in \sigma(A) $, and thus $ s(\phi^\star, \phi^a) = \sum_{\lambda} \delta_\lambda(\cS^a) $ counts the number of eigenspace directions that remain unobservable under the approximate security measure $ \phi^a $ but would be observable under the optimal one.
	In particular, $ s(\phi^\star, \phi^a)  = 0 $ if and only if the optimal and approximate security measures achieve covering the same eigenmodes.
	Hence, $ s(\phi^\star, \phi^a) $ provides a performance metric for assessing the approximation performance of Algorithm \ref{algorithm:approximate}.

	The following proposition provides an upper bound on the approximation performance.
	\begin{proposition}
		\label{proposition:approximate}
		Suppose $ \alpha_{\phi^\star}(A,C) = + \infty $. Then, we have
		\begin{align}
			\label{eq:approximate_upperbound}
			s(\phi^\star, \phi^a) \leq \frac{1}{\mathrm{e}}\sum_{\lambda \in \sigma(A)} r_\lambda,
		\end{align}
		where $ \mathrm{e} $ is the base of the natural logarithm.
	\end{proposition}
	\begin{proof}
		From (\ref{eq:delta_lambda}) and (\ref{eq:R_S}), for $ \cS \subseteq \V $, we have $ R(\cS) = \sum_{\lambda \in \sigma(A)} r_\lambda - \sum_{\lambda \in \sigma(A)} \delta_\lambda(\cS) $, which implies that, for some $ \cS_1, \cS_2 \subseteq \V $, the following holds:
		\begin{align*}
				\sum_{\lambda \in \sigma(A)}\left(\delta_\lambda(\cS_1)-\delta_\lambda(\cS_2)\right) = R(\cS_2) - R(\cS_1).
		\end{align*}
		Hence, we obtain $ s(\phi^\star, \phi^a) = R(\cS^\star) - R(\cS^a) $.
		The premise $ \alpha_{\phi^\star}(A,C) = + \infty $ implies $ \delta_\lambda(\cS^\star) = 0 $ for all $ \lambda \in \sigma(A) $, which is equivalent to that $ R(\cS^\star) $ is maximized under the budget constraint as $ R(\cS^\star) = \sum_{\lambda \in \sigma(A)} r_\lambda $.
		Recalling $ R(\cS) $ is submodular and nondecreasing, then, the approximation analysis in \cite[Theorem 1]{Sviridenko} can be directly utilized.
		More specifically, the worst-case performance guarantee of Algorithm~\ref{algorithm:approximate} is $ 1 - \mathrm{e}^{-1} $.
		Hence, we have
		\begin{align*}
			R(\cS^a) \geq \left(1-\frac{1}{\mathrm{e}}\right)\max_{\cS\subseteq \V, c(\cS)\leq \beta} R(\cS) = \left(1-\frac{1}{\mathrm{e}}\right)R(\cS^\star).
		\end{align*}
	Therefore, it follows that
	\begin{align*}
		 s(\phi^\star, \phi^a) = R(\cS^\star) -R(\cS^a) \leq \frac{1}{\mathrm{e}} R(\cS^\star)= \frac{1}{\mathrm{e}}\!\sum_{\lambda \in \sigma(A)} r_\lambda.
	\end{align*}
	\end{proof}
	
	While $ P_0 $ is originally formulated in terms of the security index, this objective is difficult to approximate directly.
	Instead, we analyze Algorithm~\ref{algorithm:approximate} through the surrogate objective $ R(\cS) $, which quantifies eigenmode coverage.
	The upper bound (\ref{eq:approximate_upperbound}) shows that, in the worst case, the approximate measure $ \phi^a $ can leave at most $ 1/\mathrm{e} $ fraction of the total eigenspace dimensions uncovered, compared with the optimal one $ \phi^\star $.
	In other words, Algorithm~\ref{algorithm:approximate} achieves at least $ 1-1/\mathrm{e} $ fraction of the optimal eigenmode coverage while requiring only polynomial-time computation.
	Therefore, although Algorithm~\ref{algorithm:approximate} is an approximation algorithm, it still guarantees a constant-factor performance bound with respect to the metric $ s(\phi^\star, \phi^a) $.

	\section{Distributed Resilient Estimation and Control with Optimal Security Measure}
	\label{section:distributed_algorithm}
	In this section, we deal with the first problem presented in Subsection \ref{subsection:problems}, that is, we aim here to design a distributed resilient state estimator and controller, which is executed in each agent, to achieve (\ref{eq:evaluation_function}).
	We now assume that each agent implements a security measure $ \phi^\star $, which is centrally calculated by using algorithms in the previous section.
	According to Theorem \ref{theorem:security_index}, unless $ (A, C_{\cS^\star}) $ is observable, there exists an undetectable attack against the system, implying that the compromised outputs cannot be distinguished from the benign ones.
	Hence, we hereafter make the following assumption to ensure the existence of a resilient state estimator.
	\begin{assumption}
		\label{assumption:detectable}
		For a given $ \phi^\star $, $ (A, C_{\cS^\star}) $ is observable. 
	\end{assumption}
	
	In the proposed estimator and controller, each agent aims to achieve (\ref{eq:evaluation_function}) distributedly while exchanging messages with neighboring agents.
	The message exchanged from agent $ i $ to its neighboring agents at time $ k $ is given as 
	\begin{align}
		\label{eq:message}
		\mathcal{M}_i(k) = \left\lbrace u_i(k-1), \hat{x}_i(k), \hat{\xi}_i(k), \hat{\xi}_{i,\ell}(k)\right\rbrace,
	\end{align}
	where $ \hat{x}_i(k) \in \R^{n_i}$ is the estimate of $ x_i(k) $ and $ \hat{\xi}_i(k) \in \R^{n} $ is the estimate of $ x(k) $ at the $ i $th agent.
	The estimate $ \hat{x}_i(k) $ is obtained by extracting entries related to agent $ i $ from $ \hat{\xi}_i(k) $.
	This paper uses a notation of $ [\cdot]_i $ to show this extraction, that is, $ \hat{x}_i(k) = [\hat \xi_i(k) ]_i $.
	The vector $ \hat{\xi}_{i,\ell}(k) $ indicates a provisional estimate of $  \hat{\xi}_i(k) $ during an estimate consensus phase, which is provided later.
	
	\subsection{Resilient Estimator and Controller Design}
	
	\begin{algorithm}[t]
		\caption{Distributed resilient estimation and control algorithm executed at each agent $ i $ with $ \phi^\star_i $}
		\label{algorithm:estimation}
		\begin{algorithmic}[1]
			\Require $ A,B, C_i, K_i, y_i(k), \hat{\xi}_i(0), \hat{x}_i(k\!-\!1), \hat{\xi}_i (k\!-\!1),u_i(k\!-\!1), x^*_i(k), u^*_i(k),\phi^\star_i, \mathcal{M}_i,\omega,L$
			\Ensure $ \hat{x}_i(k) $, $ \hat \xi_i(k) $, and $ u_i(k) $
			\For{$ k = 1,2,\ldots $}
			\State [Input fusion] Obtain the input $ u(k-1) $ through message exchange with neighboring agents.
			\State [Time update] $ \bar \xi_i(k) = A\hat\xi_i(k-1) + Bu(k-1) $
			\State [Observation update] 
			\If{$ i \in \mathcal{S}^\star $} 	\Comment{Agent $ i $ is secure.}
			\State $ \tilde \xi_i(k) = \bar \xi_i(k)  +C_i^\top \left(y_i(k) - C_i \bar \xi_i(k)\right)$
			\Else \Comment{Agent $ i $ is normal.}
			\State $ \tilde \xi_i(k) = \bar \xi_i(k)  $
			\EndIf
			\State [Estimate consensus] $ \hat{\xi}_{i,0}(k) \leftarrow \tilde{\xi}_i (k) $
			\For{$ \ell = 1,\ldots,L $}
			\State $ \!\hat{\xi}_{i,\ell}(k)\! =\! \hat{\xi}_{i,\ell-1}(k) - \omega \!\sum_{j \in \mathcal{N}_i}\!\!\left(\!\hat{\xi}_{i,\ell-1}(k)\!-\! \hat{\xi}_{j,\ell-1}(k)\!\right)$\!
			\EndFor
			\State $ \hat \xi_i(k) \leftarrow \hat{\xi}_{i,L}(k) $ and $ \hat{x}_i(k) \leftarrow \left[\hat \xi_i(k) \right]_i$
			\State [Input update] $ u_i(k) = u^*_i(k)+K_i \left(x^*_i(k) - \hat{x}_i(k)\right) $
			\State \textbf{return} $ \hat{x}_i(k) $, $ \hat \xi_i(k) $, and $ u_i(k) $
			\EndFor
		\end{algorithmic}
	\end{algorithm}
	
	The proposed distributed resilient estimation and control algorithm with the security measure $ \phi^\star $, which is executed at each agent, can be summarized as Algorithm~\ref{algorithm:estimation}.
	We derive a multi-time-scale estimator with four phases, i.e., input fusion, time update, observation update, and estimate consensus.
	The input fusion phase aims to have each agent collect the previous time input of all other agents.
	To this end, using an information fusion algorithm such as a simple consensus algorithm (see, e.g., \cite{DistributedAlgorithms}), multiple information exchanges of $ u_i(k-1) $ are performed between each adjacent agent.
	In the time update phase, each agent computes \textit{a priori} estimate based on the overall input:
	\begin{align}
		\label{eq:time_update}
		\bar \xi_i(k) = A\hat\xi_i(k-1) + Bu(k-1).
	\end{align}
	In the observation update phase, using the sensor output, the following estimate is obtained:
	\begin{align}
		\label{eq:observation_update}
		\tilde \xi_i(k) \! = \!\left\lbrace \!\!\begin{array}{ll}
			\bar \xi_i(k)  \!+\!C_i^\top \left(y_i(k) - C_i \bar \xi_i(k)\right), &\!\!\mathrm{if}~\phi^\star_i = \sS,\\
			\bar \xi_i(k), &\!\!\mathrm{if}~\phi^\star_i = \sN.\end{array} \right. \!\!
	\end{align}
	This observation update can be interpreted as follows:
	Only the secure agents utilize the sensor output because the sensor measurement can be trusted, while the normal ones do not use its sensor output because it may contain a malicious value.
	
	In the estimate consensus phase, each agent communicates with its neighbors for $ L \geq 1 $ times.
	For $ \ell = 1,\ldots, L $ and $ \omega > 0 $, we have
	\begin{align}
		\label{eq:estimate_consensus}
		\hat{\xi}_{i,\ell}(k) = \hat{\xi}_{i,\ell-1}(k) - \omega \sum_{j \in \mathcal{N}_i}\left(\hat{\xi}_{i,\ell-1}(k)- \hat{\xi}_{j,\ell-1}(k)\right)
	\end{align}
	with $ \hat{\xi}_{i,0}(k) = \tilde \xi_i(k) $ and $ \hat{\xi}_i(k) = \hat{\xi}_{i,L}(k)$.
	The design of the communication rate $ L $ and consensus gain $ \omega $ will be discussed in the next subsection.
	In the $ \ell $th communication of the estimate consensus phase, agent $ i $ transmits its estimate $ \hat{\xi}_{i,\ell-1}(k) $ to its neighbors.
	As in (\ref{eq:observation_update}), since normal agents do not exploit the sensor output, it is vital to determine the communication rate $ L $ and the consensus gain $ \omega  $ to guarantee the bounded estimation error.
	The estimate $ \hat{x}_i(k)  $ is finally obtained as $ \hat{x}_i(k) = [\hat \xi_i(k) ]_i $. 
	
	Based on the state estimate of each agent, we next design an observer-based distributed controller as follows:
	\begin{align*}
		u_i(k) = u^*_i(k)+K_i \left(x^*_i(k) - \hat{x}_i(k)\right),
	\end{align*}
	where $ u^*_i(k) \in \R^{m_i}$ is the desired input, $ x^*_i(k) \in \R^{n_i}$ is the desired state of the $ i $th agent at time $ k $, and $ K_i \in \R^{m_i \times n_i} $ is the gain parameter of agent $ i $ to be determined.
	We assume that the desired state of each agent follows the nominal dynamics: $ x_i^*(k+1) = A_ix^*_i(k) + B_i u^*_i(k)$.

	\subsection{Performance Analysis}
	In this subsection, we show performance properties of Algorithm~\ref{algorithm:estimation}.
	We first obtain the following theorem describing an upper bound on the estimation error.
	
	\begin{theorem}
		\label{theorem:estimator_performance}
		Denote
		\begin{align}
			F &\!\triangleq\! \left(I_n - \frac{1}{N}\sum_{i \in \mathcal{S}^\star }C_i^\top C_i\right)A, \\
			\theta_0 & \!\triangleq\! 	\max \left\lbrace 1,\max_{i \in \cS^\star}\left\| I_{n} - C_i^\top C_i\right\| \right\rbrace,~\gamma_\bot\! \triangleq\!\dfrac{\lambda_{\max}^\cL- \lambda_2^\cL}{\lambda_{\max}^\cL+ \lambda_2^\cL}.		
	\end{align}
		Under Assumption~\ref{assumption:detectable}, consider Algorithm~\ref{algorithm:estimation} with the optimal security measure $ \phi^\star $ and $ \omega$ as
		\begin{align}
			\label{eq:omega}
			\omega = \dfrac{2}{\lambda_{2}^\cL + \lambda_{\max}^\cL}.
		\end{align}
		If the following statements hold:
		\begin{align}
			\label{eq:L}
		\dfrac{\ln \left(\theta_0 \| A\|\right)}{\ln \gamma_{\bot}^{-1}} &< L,\\
			\label{eq:F}
			\| F\| &< 1, \\
			\label{eq:last_condition}
			\gamma_{\bot}^L \theta_0^2 \|A\|^2 & < \left(1-\gamma_\bot^L \theta_0 \|A\|\right)\left(1-\|F\|\right),
		\end{align}
		then, for all $ i \in \mathcal{V} $, the estimation error $ e_i(k) \triangleq \hat{x}_i(k) - x_i(k) \in \R^{n_i}$ follows
		\begin{align}
			\label{eq:error}
			\limsup_{k \rightarrow \infty} \left\|e_i(k) \right\| \leq \dfrac{D\left(1+\gamma_\bot^L\left(1-\|F\|+\theta_0 \|A\|\right)\right)}{\zeta},
		\end{align}
		where 
		\begin{align*}
			D & \triangleq \theta_0 \sqrt{N}\delta_w + \nu_0 \delta_v,~~\nu_0 \triangleq \max_{i \in \cS^\star}\left\| C_i\right\|,\\
			\zeta & \triangleq \left(1-\gamma_\bot^L \theta_0 \|A\|\right)\left(1-\|F\|\right)-\gamma_{\bot}^L \theta_0^2 \|A\|^2.
		\end{align*}
	\end{theorem}
	\begin{proof}
		See Appendix~\ref{appendix:proof_theorem2}.
	\end{proof}
	
	This theorem shows that, under (\ref{eq:L})--(\ref{eq:last_condition}), our proposed distributed algorithm ensures bounded estimation errors in each agent.
	Condition (\ref{eq:L}) specifies a lower bound on the number of communications among agents.
	Even if $ A $ is not stable and $ \| A \| > 1 $, (\ref{eq:L}) holds for a sufficiently large number of communication rounds $ L $.
	Also, (\ref{eq:F}) holds even when $ \| A \| > 1 $.
	Intuitively, as more agents are secured, the norm of $ I_n - \frac{1}{N}\sum_{i \in \mathcal{S}^\star }C_i^\top C_i  $ tends to decrease.
	Consequently, the condition $ \|F\| < 1 $ is easier to satisfy, which shows that implementing security measures contributes to the stability of the distributed state estimation under sensor attacks.
	Condition (\ref{eq:last_condition}) can be viewed as a small-gain condition: it compares the interaction term $ 	\gamma_{\bot}^L \theta_0^2 \|A\|^2  $ with the stability margins $ 1-\gamma_\bot^L \theta_0 \|A\| $ and $1-\|F\| $ of the consensus and secure observer dynamics, and requires this interaction to be weak enough so that the estimator remains stable.
	Additionally, the upper bound condition (\ref{eq:error}) can be interpreted as follows:
	Decreasing the upper bound on noises $ D $ reduces the estimation error.
	Increasing $ L $ makes $ \gamma_\bot^L $ smaller (because $ 0 < \gamma_\bot < 1 $), which contributes to decreasing the estimation error.
	Implementing security measures for more agents tends to decrease the norm of $ F $, thereby reducing the estimation error.

	In the following proposition, we show the control performance of Algorithm~\ref{algorithm:estimation}.
	\begin{proposition}
		\label{proposition:control}
		Suppose that Assumption~\ref{assumption:detectable} holds.
		Considering Algorithm~\ref{algorithm:estimation} with the optimal security measure $ \phi^\star $ and $ \omega $ as (\ref{eq:omega}), if (\ref{eq:L})--(\ref{eq:last_condition}) hold and $ K_i $ satisfies
		\begin{align}
			\label{eq:K_p}
			\rho\left(A_i - B_i K_i\right) < 1,~\forall i \in \V,
		\end{align}
		then, for all $ i \in \V $, the control error $ \tilde e_i(k) \triangleq x_i(k) - x^*_i(k) \in \R^{n_i} $ follows
		\begin{align}
			\label{eq:error_control}
			\limsup_{k \rightarrow \infty} \left\| \tilde e_i (k) \right\| \leq &\left(\sum_{t=0}^{\infty} \left\| (A_i - B_i K_i)^t \right\|\right) \nonumber\\
			&~~~~~~\cdot \left(\left\| B_iK_i\right\|\Delta_e^\infty + \delta^w_i\right),
		\end{align}
		where $ \Delta_e^\infty  $ is the asymptotic upper bound of the estimation error provided by (\ref{eq:error}).
	\end{proposition}
	\begin{proof}
		See Appendix~\ref{appendix:proof_proposition1}.	
	\end{proof}
	
	Note that $ \sum_{t=0}^{\infty} \left\| (A_i - B_i K_i)^t \right\| $ is bounded when $ \rho(A_i - B_i K_i)< 1  $.
	Thus, we see that the control error is also asymptotically upper bounded for all agents by designing the feedback gain $ K_i $ as (\ref{eq:K_p}).
	Finally, we have the following corollary to provide a solution to the first problem of Subsection~\ref{subsection:problems}.
	\begin{corollary}
		\label{cororally:solution}
		Suppose that Assumption~\ref{assumption:detectable} holds.
		Considering Algorithm~\ref{algorithm:estimation} with the optimal security measure $ \phi^\star $ and $ \omega $ as (\ref{eq:omega}), if (\ref{eq:L})--(\ref{eq:last_condition}) hold and $ K_i $ satisfies (\ref{eq:K_p}), then (\ref{eq:evaluation_function}) holds for some $ \Delta > 0 $.
		In particular, if the system is noiseless, i.e., $ \delta_w = \delta_v  = 0$, we obtain
		\vspace{-2mm}
		\begin{align*}
			\limsup_{k \rightarrow \infty} \dfrac{1}{N} \sum_{i=1}^{N} \left(\left\| \hat{x}_i(k) - x_i(k)\right\|  + \left\| x_i(k) - x^*_i(k)\right\|\right) = 0.
		\end{align*}
	\end{corollary}
	\begin{proof}
		The proof follows from Theorem~\ref{theorem:estimator_performance} and Proposition~\ref{proposition:control}.
	\end{proof}
	
	%
	%
	%

	\section{Numerical Simulations with Vehicle Platooning}
	\label{section:simulation}
	In this section, we provide numerical simulations to corroborate the effectiveness of the proposed framework.
	We use a vehicle platooning system as the simulation model. 
	
	\subsection{Platoon Model}
	The aim of platooning is to control the speeds of all vehicles to a desired value while maintaining a safe distance between any two adjacent vehicles.
	Our example employs the following linearized third-order longitudinal vehicle dynamics \cite{Platooning-01,Platooning-02,Platooning-03}:
	\begin{align}
		\left\lbrace \begin{array}{l}
			\dot{\mathsf{p}}^v_{i}  = \mathsf{v}^v_{i}, \\
			\dot{\mathsf{v}}^v_{i}  = \mathsf{a}^v_{i}, \\
			\dot{\mathsf{a}}^v_i  = -\tau_i^{-1}\mathsf{a}^v_{i} + \tau_i^{-1}u_i,
		\end{array}\right.~i \in \{1,\ldots, N\},
	\end{align}
	where $ \mathsf{p}^v_{i} \in \R $, $ \mathsf{v}^v_{i} \in \R $, and $\mathsf{a}^v_{i} \in \R $ denote the position, velocity, and acceleration, respectively, $ \tau_i \in \R $ represents the inertia time constant, and $ u_i \in \R $ is the commanded acceleration of the $ i $th vehicle.
	$ N $ denotes the total number of vehicles.
	In general, different types of vehicles have different inertia parameters $ \tau_i $, which results in a heterogeneous vehicle model.
	This third-order heterogeneous model captures vehicle heterogeneity and inertia-delay characteristics in longitudinal dynamics, and is therefore widely used in the literature as a basis for analysis.
	In this paper, we use a discretized model with a sampling time of $ T_s = 0.01~\mathrm{sec} $ and assume $ \delta^w_i = 0.1 $ as the bound of the process noise.
	The discretized state vector of vehicle $ i $ can be characterized as $ x_i(k) = \left(\mathsf{p}^v_{i}(k), \mathsf{v}^v_{i}(k), \mathsf{a}^v_{i}(k)\right)^\top \in \R^3 $.
	
	Each vehicle is able to obtain its position and velocity measurements through a GPS receiver.
	Also, except for the first vehicle, each vehicle $ i $ can get the relative position and velocity measurements with respect to its front vehicle (i.e., vehicle $ i-1 $) through a sensor.
	Hence, the measurement of each (normal) vehicle is given as
	\begin{align*}
		y_i(k) \!=\!\! \left\lbrace \!\! \begin{array}{ll}
			\left[\!\!\begin{array}{c}
				\mathsf{p}^v_{i}(k) \\ \mathsf{v}^v_{i}(k)
			\end{array}\!\!\right] + a_i(k) + v_i(k), &\!\!\mathrm{if}~i = 1, \\
			\left[\!\!\begin{array}{c}
				\mathsf{p}^v_{i}(k) \\ \mathsf{v}^v_{i}(k) \\
				\mathsf{p}^v_{i}(k) - \mathsf{p}^v_{i-1}(k) \\ \mathsf{v}^v_{i}(k) - \mathsf{v}^v_{i-1}(k)
			\end{array}\!\!\right] + a_i(k) + v_i(k), &\!\! \mathrm{otherwise},
		\end{array} \right.\!\!
	\end{align*} 
	where $y_1(k) \in \R^2 $ and $ y_i(k) \in \R^4 $ for $ i \in \{2,\ldots,N\} $.
	Here, the attack $a_i(k)$ and noise $v_i(k)$ have the same dimensions as the output.
	In this simulation, we set $ \delta^v_i = 0.1 $ as the bound on the measurement noise.
	The measurement matrix $ C_i $ is obtained through this observation model.
	Regarding the cost of the security measure, we suppose that $ c^\mathcal{N}_i = 1 $ and $c^\mathcal{S}_i = 30 $ for all vehicles.

	\subsection{Security Implementation and Approximation Performance}

\begin{figure}[t]
		\begin{center}
			\includegraphics[width=\linewidth]{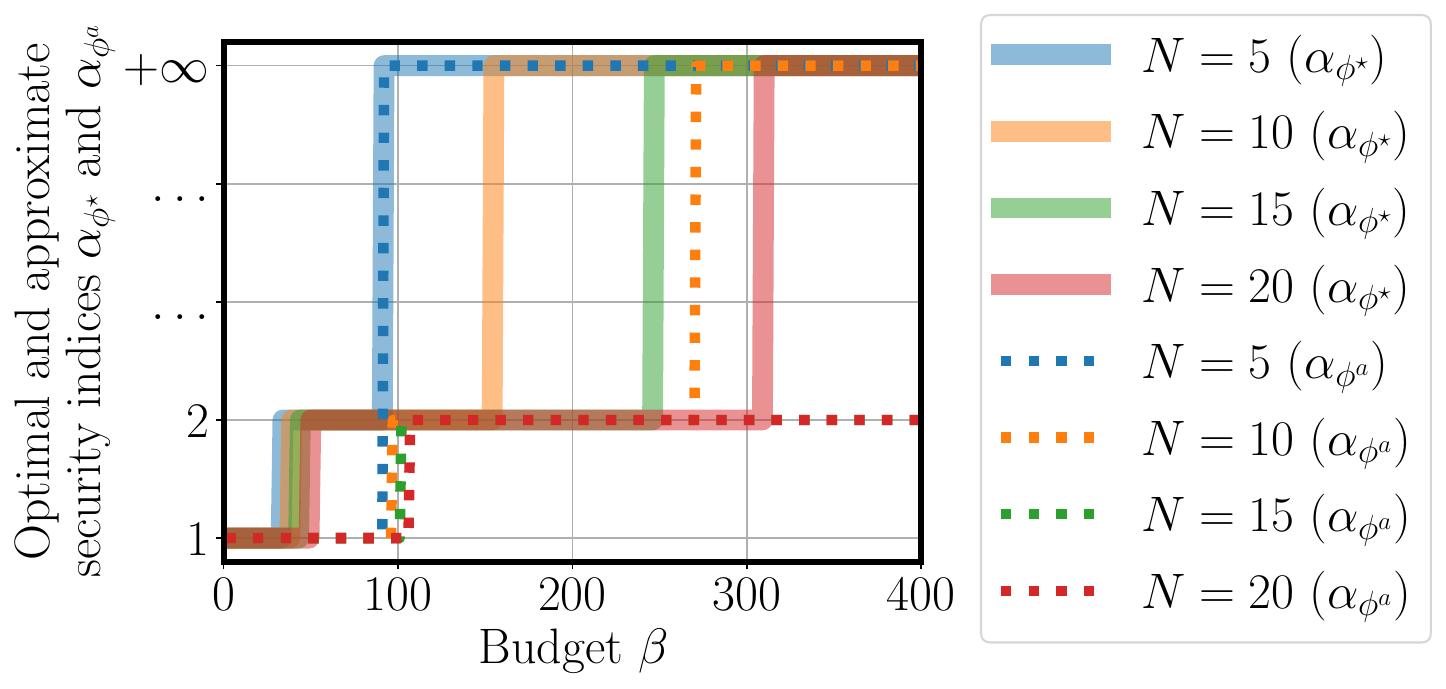}
			\vspace{-6mm}
			\caption{The budget $ \beta $ versus the optimal and approximate security indices $ \alpha_{\phi^\star}(A,C) $ and $ \alpha_{\phi^a}(A,C) $ for different $ N $. 
					Bold solid lines indicate optimal security index and dotted lines indicate the approximate one.
					The value $ +\infty $ indicates that there is no undetectable attack against the system with implemented security measures.}
			\label{fig:alpha_beta}
			
			\vspace{2mm}
			
			\includegraphics[width=0.9\linewidth]{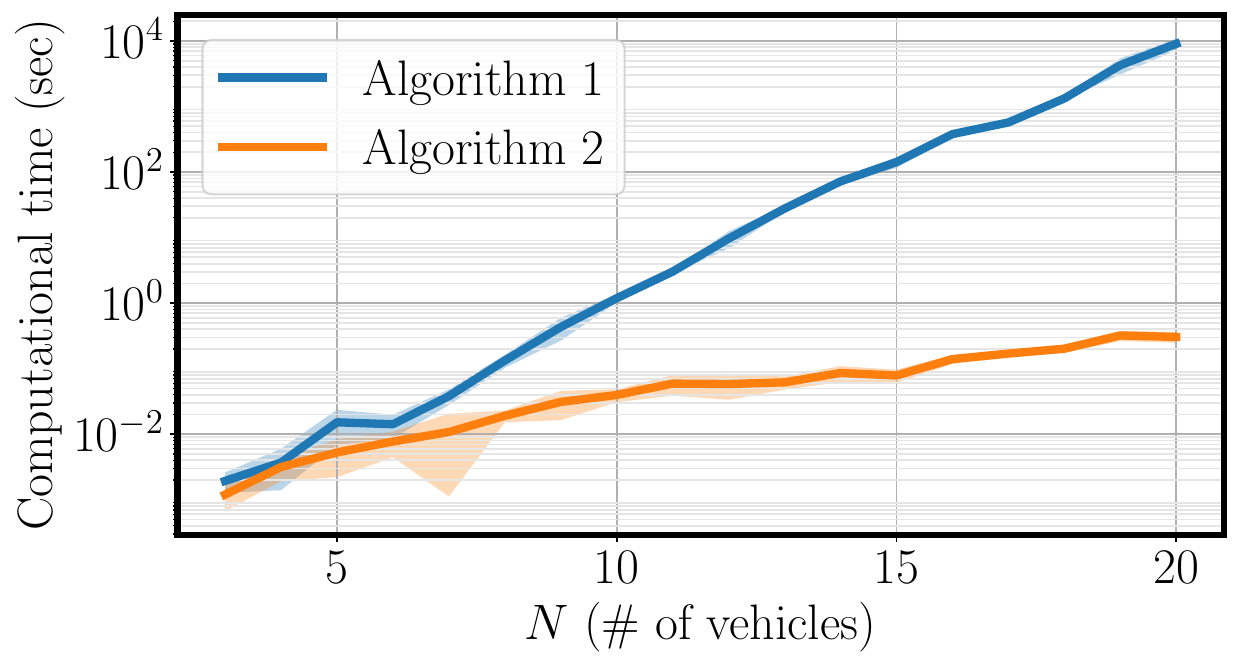}
			\vspace{-2.5mm}
			\caption{Computational-time comparison between Algorithms~\ref{algorithm:centralized} and \ref{algorithm:approximate} for different $ N $ and the budget $ \beta = 30N $.
				Shaded areas represent $ \pm $ one standard deviation over $ 10 $ runs.}
			\label{fig:computational_time}
		\end{center}
		\vspace{-7mm}
	\end{figure}

	We first present optimal security implementation results.	
	Fig.~\ref{fig:alpha_beta} shows the relationship between the budget $ \beta $ and the optimal and approximate security indices $ \alpha_{\phi^\star}(A,C) $ and $ \alpha_{\phi^a}(A,C) $ based on Algorithms \ref{algorithm:centralized} and \ref{algorithm:approximate}, respectively, for different numbers of vehicles $ N $.
	Here, the inertia time constant $ \tau_i $ is randomly chosen from $ 0.5 $ to $ 1 $.
	
	First, consider the optimal security index $ \alpha_{\phi^\star}(A,C) $.
	The security index of the original system (without any security implementation) is 1, since the overall system loses its observability if the output of the $N$th vehicle (i.e., the last vehicle) is compromised.
	If we have a budget such that the last vehicle implements a security measure, namely $ \beta \geq c_i^\mathcal{S} +(N-1)c_i^\mathcal{N} $, then the optimal security index increases to 2.
	Furthermore, if a larger budget is available, the system operator can achieve the maximum security index $ \alpha_{\phi^\star} = + \infty $, where the optimal security measure is given as 
	\begin{align*}
		\phi^\star = \left\lbrace \begin{array}{ll}
			\left[\sS, \sN, \sS, \sN, \ldots\right]^\top, & \mathrm{if}~N~\mathrm{is~odd}, \\
			\left[\sN, \sS, \sN,\sS, \ldots\right]^\top, & \mathrm{if}~N~\mathrm{is~even},
		\end{array} \right.
	\end{align*}
	since by installing security measures on every second vehicle, the measurement information of both the secure vehicle and the vehicle in front of it will be protected.
	In this case, there are no undetectable attacks for the system.	
	
	We next consider the approximation performance of the approximate security index $ \alpha_{\phi^a}(A,C) $ obtained from Algorithm~\ref{algorithm:approximate}, as illustrated in Fig.~\ref{fig:alpha_beta}.
	For large budgets, $\alpha_{\phi^a}(A,C)$ coincides with the optimal security index $\alpha_{\phi^\star}(A,C)$, whereas a  gap appears for small budgets.
	This behavior is consistent with Proposition~\ref{proposition:approximate}, which guarantees that Algorithm~\ref{algorithm:approximate} achieves a constant-factor approximation of the optimal eigenmode coverage $R(\cS)$ for any budget.
	In particular, when the budget is large enough so that full eigenmode coverage is achievable, any secure set with full coverage yields the maximal security index by Lemma~\ref{lemma:undetectable}.
	In our example, Algorithm~\ref{algorithm:approximate} attains full coverage once the budget exceeds this threshold, and
	thus also achieves the maximal security index, as observed in Fig.~\ref{fig:alpha_beta}.
	On the other hand, in low-budget scenarios, the algorithm still maximizes $R(\cS)$ only approximately, and this surrogate objective does not necessarily lead to the optimal security index. 
	This explains the gap between $\alpha_{\phi^\star}(A,C)$ and $\alpha_{\phi^a}(A,C)$ in our example.

	\subsection{Computation Performance}
	
	We then compare the computational performance\footnote{The simulations are performed on a laptop equipped with an Intel Core i7-1165G7 2.80GHz and a 16 GB memory chip.} of Algorithms~\ref{algorithm:centralized} and \ref{algorithm:approximate} for different numbers of vehicles $ N $ with the budget $ \beta = 30N $.
	The computational-time comparisons in these two algorithms are given in Fig. \ref{fig:computational_time}.
	From this figure, it can be observed that Algorithm~\ref{algorithm:approximate} outperforms Algorithm~\ref{algorithm:centralized} in terms of the computational cost.
	Therefore, in large-budget scenarios, it is reasonable to use Algorithm~\ref{algorithm:approximate} to efficiently compute an approximate security index, which coincides with the optimal one in the present example.
	
	\subsection{Estimation and Control Performance}
	\begin{figure}[t]
		\begin{center}
			\vspace{-3mm}
			\includegraphics[width=0.95\linewidth]{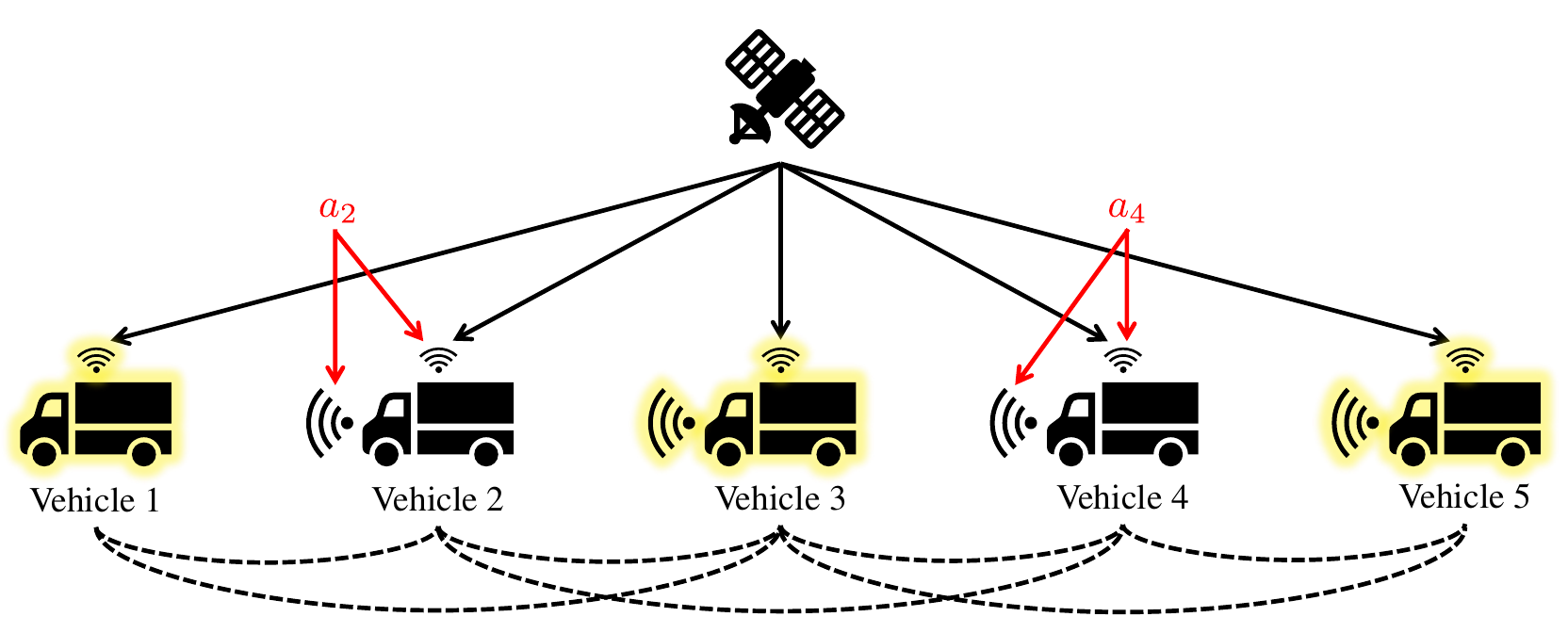}
			\vspace{-1mm}
			\caption{Model of a platoon of five vehicles (agents). The measurements of vehicles 2 and 4 are compromised by a malicious adversary. Vehicles 1, 3, and 5 are designated as secure by the optimal security measure $ \phi^\star $. The communication topology among vehicles is shown by the dashed lines at the bottom and coincides with that in Fig.~\ref{fig:system_diagram_overall}.}
			\label{fig:simulation_vehicle}
		\end{center}
		\vspace{-3mm}
	\end{figure}

	We next examine the estimation and control performance of the proposed algorithm (Algorithm~\ref{algorithm:estimation}) in the presence of sensor attacks.
	We here consider the five-vehicle platooning (i.e., $ N = 5 $), as depicted in Fig.~\ref{fig:simulation_vehicle}.
	Assuming $ \beta = 150 $, we obtain $ \phi^\star =\left[\sS, \sN, \sS, \sN, \sS\right]^\top $, i.e., vehicles 1, 3, and 5 are equipped with security measures against malicious attacks, which maximizes the security index of the entire system.
	For the inertia parameter of each vehicle, we set $ \tau_1 = 0.8, \tau_2 = 0.65, \tau_3 =0.5, \tau_4 = 0.4, \tau_5 = 0.3$.
	For the initial state of each vehicle, we set $ x_1(0) = (200,10,0)^\top$, $x_2(0)=(100,8,0)^\top$, $x_3(0)=(50,6,0)^\top$, $x_4(0)=(20,4,0)^\top$, $x_5(0)=(0,2,0)^\top $.
	The overall dynamics of all five vehicles are given by (\ref{eq:overall_dynamics}), where $ x(k) = (x_1(k)^\top,\ldots,x_5(k)^\top)^\top \in \R^{15}$.
	We assume that the 2nd and 4th vehicles are subject to malicious sensor attacks.
	The designed attack injections for each vehicle are given as $ a_2(k) = -C_2 x(k) - v_2(k) $ and $ a_4(k) =  -C_4 x(k)- v_4(k)$ for all $ k $, which make the compromised outputs identically zero.
	
	
	For the communication model, we assume that each vehicle can communicate with vehicles within two positions, namely, $ \mathcal{N}_1 = \{2, 3\}$, $\mathcal{N}_2 = \{1,3,4\}$, $\mathcal{N}_3 = \{1,2,4,5\}$, $\mathcal{N}_4=\{2,3,5\}$, $\mathcal{N}_5 = \{3,4\}. $
	
	
	\begin{figure}[t]
		\centering
		\includegraphics[width=\linewidth]{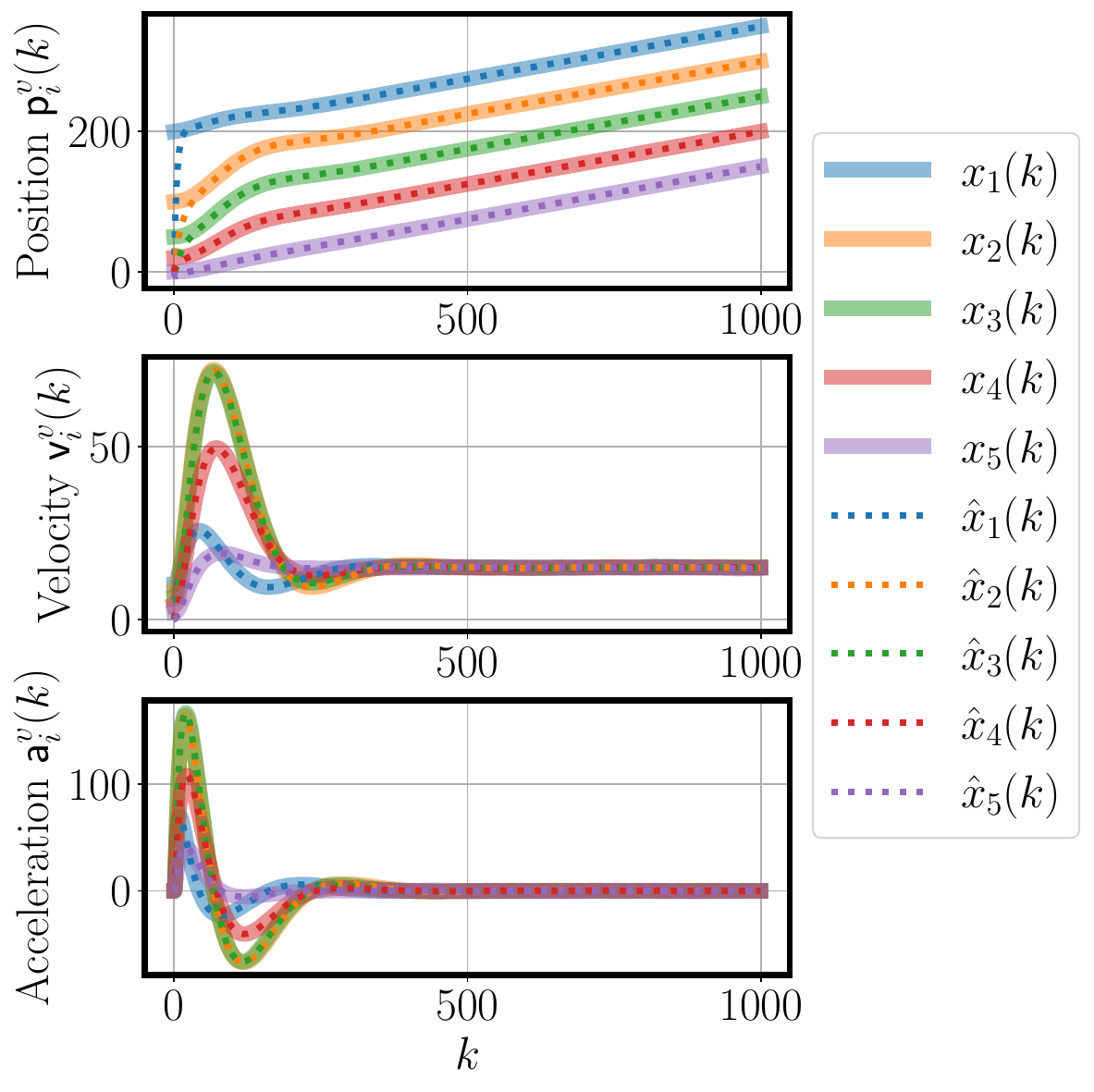}
		\vspace{-6.5mm}
		\caption{Performance of the proposed estimation and control algorithm (Algorithm~\ref{algorithm:estimation}). Bold solid lines indicate true states, and dotted lines indicate state estimates.}
		\label{fig:simulation_result}
		\vspace{-4mm}
	\end{figure}
	
	With $ \omega = 2/\left(\lambda_{2}^\cL + \lambda_{\max}^\cL \right) = 0.3037$, $ L = 5 $ and gains $ K_i $ satisfying (\ref{eq:K_p}) for Algorithm~\ref{algorithm:estimation}, the estimation and control results in each vehicle are provided in Fig.~\ref{fig:simulation_result}.
	Here, the desired state of each vehicle is assumed to follow the nominal dynamics with zero input, i.e., $ x^*_i(k+1)=A_ix^*_i(k) $ with each initial state $ x^*_1(0) = (200,15,0)^\top $, $ x^*_2(0) = (150,15,0)^\top $, $ x^*_3(0) = (100,15,0)^\top $, $ x^*_4(0) = (50,15,0)^\top $, and $ x^*_5(0) = (0,15,0)^\top $.
	From Fig.~\ref{fig:simulation_result}, it can be observed that the platooning control achieves the desired formation in a distributed manner, and these states are well estimated by using the proposed algorithm, even under attacks on two agents.

	\begin{figure*}[t]
		\centering
		\includegraphics[width=0.8\linewidth]{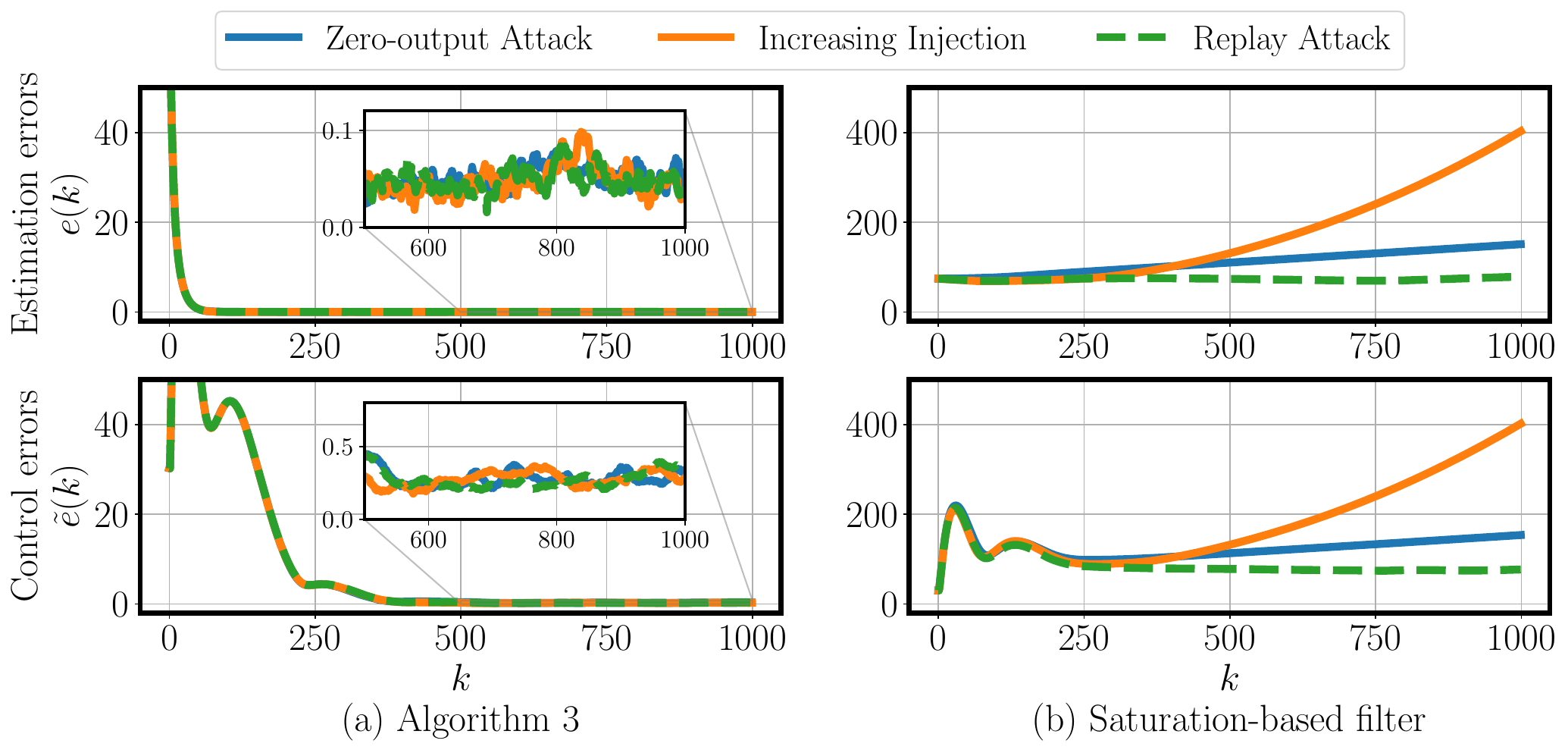}
		\vspace{-2mm}
		\caption{Estimation error $ e(k) $ and control error $ \tilde e(k) $ of Algorithm~\ref{algorithm:estimation} and the distributed saturation-based filter \cite{2022TACJohanssson} under the three attack scenarios.}
		\label{fig:error}
		\vspace{-3mm}
	\end{figure*}


	%
	
	Moreover, we compare our algorithm with the distributed saturation-based filter \cite{2022TACJohanssson} under three typical attack scenarios targeting the 2nd and 4th vehicles, including the zero-output attack introduced previously, increasing-signal attack $ a_i(k) = 10k\cdot \bm{1}_{p_i} $, and replay attack $ a_i(k) = -C_i x(k) - v_i(k) + y_i(k-50)$ for $ k\geq 51 $, where $ y_i(k) $ is the true measurement at time $ k $.
	The estimation and control errors of our algorithm and the distributed saturation-based filter \cite{2022TACJohanssson} are shown in Fig.~\ref{fig:error}, where the global estimation error $ e(k) $ and control error $ \tilde e(k) $ are defined as $ e(k) \triangleq \frac{1}{N}\sum_{i=1}^N \left\| e_i(k)\right\| $ and $ \tilde e(k) \triangleq \frac{1}{N}\sum_{i=1}^N \left\| \tilde e_i(k)\right\| $, respectively.
	The saturation threshold for the distributed saturation-based filter is set as $ 1 $ (for details, see \cite{2022TACJohanssson}) and all other parameters are the same as above.
	When the sensors of vehicles 2 and 4 are compromised, the boundedness condition of the algorithm in \cite{2022TACJohanssson} is violated, and hence, the estimation and control errors are divergent in all attack scenarios, as shown on the right-hand side of Fig.~\ref{fig:error}.
	On the other hand, even in these adversarial environments, the proposed algorithm achieves bounded estimation and control errors, which illustrates the effectiveness of implementing the security measure and the proposed method.
	
	Note that the numerical examples in this paper are chosen so that all inter-vehicle distances remain positive, as we focus on demonstrating the proposed resilient estimation and control against sensor attacks.
	In practice, the proposed framework can be combined with collision-avoidance methods (e.g., using a safety-aware predictive control scheme or a barrier function) to ensure collision-free operation.
	
	\section{Conclusion and Future Work}
	\label{section:conclusion}
	This paper studied a distributed resilient state estimation and control problem against malicious sensor attacks, considering the implementation of security measures for the system.
	We first showed that the system's resilience, characterized by the security index, is maximized by employing a security measure that renders the resulting system observable.
	This implies that any knowledgeable and capable attacker cannot perform undetectable attacks.
	Then, we investigated an algorithm to implement the optimal security measure strategically.
	In general, this calculation is computationally intensive, but we also provided a computationally efficient approximation algorithm.
	We proposed a distributed resilient state estimation and control framework based on the optimal security measure, and analyzed the boundedness conditions of the proposed scheme.
	Finally, the proposed framework was applied to vehicle platooning in numerical simulations.
	
	An important limitation of the present formulation is that the distributed algorithm assumes protected inter-agent communication.
	Future work includes extending the strategic and distributed security design to account for communication-link attacks or Byzantine agents, possibly by combining the present observability-based design with trusted-node or	Byzantine-resilient estimation frameworks such as \cite{2021TCNSSundaram}.
	Additionally, we are interested in a scenario in which Assumption~\ref{assumption:detectable} does not hold (i.e., the security index is not maximized).
	\appendices

	\section{Proof of Theorem~\ref{theorem:security_index}}
	\label{appendix:proof_theorem1}
	In this proof, we utilize a property of the observability of $ (A,C_\cS) $ for a set $ \cS \subseteq \V $.
	Referring to the original definition of the system observability (see, e.g., \cite{1990Sontag}), the pair $ (A,C_\cS) $ is observable if and only if the linear map $ \mathcal{O}_\cS $ defined by $ \mathcal{O}_\cS \triangleq [\begin{array}{cccc}
				C_{\cS}^\top  & \left(C_{\cS} A \right)^\top & \cdots & \left(C_{\cS}  A^{n-1} \right)^\top
	\end{array}]^\top $ satisfies $ \ker \mathcal{O}_\cS = \{0\}$.
	
	For the sufficiency, we resort to the contradiction.
	Assume that $ (A, C_\cS) $ is observable, but an attacker is able to design undetectable attacks.
	Then, there exists an undetectable attack sequence $ a(k) $ such that $ y(x^1(0), u, a, \phi, k) = y(x^2(0), u, 0, \phi, k) $ for two distinct initial states $ x^1(0) \in \R^n $ and $ x^2(0) \in \R^n $.
	Defining two different state trajectories from $ x^1(0) $ and $ x^2(0) $ as $ x^1(k) $ and $ x^2(k) $, respectively, since the measurements due to these trajectories and the undetectable attack coincide, we have $ C x^1(k) + v(k) + V(\phi)a(k) = Cx^2(k) + v(k). $
	It follows that $ \blocksupp{V(\phi)a(k)} \subseteq \N = \V \backslash \cS $, which indicates that $ C_\cS (x^2(k) -x^1(k)) =0 $ for any $ k $. 
	The trajectories $ x^1(k) $ and $ x^2(k) $ can be represented as
	\begin{align*}
		x^1(k) &\!= \!A^k x^1(0) \!+ \!H_{u,k}\underbrace{\left[\!\!\begin{array}{c}
				u(0) \\ \vdots \\ u(k-1)
			\end{array}\!\!\right]}_{U(k)}\!+\!H_{w,k}\underbrace{\left[\!\!\begin{array}{c}
				w(0) \\ \vdots \\ w(k-1)
			\end{array}\!\!\right]}_{W(k)}, \\
		x^2(k) &\!=\! A^k x^2(0) + H_{u,k}U(k)\!+\!H_{w,k}W(k),
	\end{align*}
	where $ H_{u,k} $ is the input-to-state matrix calculated by $ A $ and $ B $ and $ H_{w,k} $ is the noise-to-state matrix similarly obtained from $ A $ and $ I_n $.
	Therefore, it holds that $ C_\cS (x^2(k) -x^1(k)) =C_\cS A^k (x^2(0)-x^1(0))=0 $ for any $ k $.
	This implies that the nonzero vector $ x^2(0) - x^1(0) $ satisfies $ x^2(0)-x^1(0) \in \ker \mathcal{O}_\cS $, which contradicts the assumption that $ (A, C_\cS) $ is observable.
	
	For the necessity, then, we prove it by using contraposition.
	To this end, assume that $ (A, C_\cS) $ is not observable for a given security measure $ \phi $, which implies that the linear map $ \mathcal{O}_\cS $ has a non-trivial kernel.
	Considering two distinct initial states $ x^1(0) \in \ker \mathcal{O}_\cS \backslash\{0\} $ and $ x^2(0) = 0$, we obtain two different trajectories as $ 	x^1(k) = A^kx^1(0)+H_{u,k}U(k)+H_{w,k}W(k)$ and $ x^2(k) = H_{u,k}U(k)\!+\!H_{w,k}W(k) $.
%
	Further, consider an attack sequence of $ a(k) = -CA^kx^1(0) $ and assume this attack is injected into the sensor output of the first state trajectory.
	Then, we have output sequences for each state trajectory as
	\begin{align*}
		y^1(k) &= C A^k x^1(0) + C H_{u,k}U(k)+ CH_{w,k}W(k) \nonumber \\
		&\hspace{40mm}+ v(k) + V(\phi)a(k), \\
		y^2(k) &= C H_{u,k}U(k) + CH_{w,k}W(k) + v(k).
	\end{align*}	
	By the relations $ \mathcal{O}_\cS x^1(0) = 0 $ and $ \blocksupp{V(\phi)a(k)} \subseteq \N = \V \backslash \cS $, it follows that $ C A^k x^1(0) + V(\phi)a(k) = C_\mathcal{N} A^k x^1(0) - C_\mathcal{N}A^kx^1(0) = 0$, which indicates that the above two output sequences coincide.
	This concludes that an attacker is able to design an undetectable attack as $ a(k) = -CA^kx^1(0) $.
	\endproof
	
	\section{Proof of Proposition \ref{proposition:undetectable_attacks}}
	\label{appendix:proof_proposition:undetectable_attacks}
	First, for $ 1) \Rightarrow 2) $, assume that there exists an undetectable attack, which implies $ (A, C_\cS) $ is not observable by Theorem~\ref{theorem:security_index}.
	Borrowing the standard definition of the system unobservability (see, e.g., \cite{1990Sontag} and \cite{Linear Systems}), it holds that $ C_\cS v = 0 $ for some nonzero $ v \in \mu(A) $.
	This implies $ v \in \ker C_\cS $, and thus it holds that $ v \in \ker C_\cS \cap \mu(A) $, which implies that
	$ \ker C_\cS \cap \E(\lambda) \neq \{0\} $ for some $ \lambda \in \sigma(A) $.

%
%
%
%
	For $ 2) \Rightarrow 3) $, assume there is a nonzero vector $ v \in \ker C_\cS \cap \mathcal{E}(\lambda) $ for some $ \lambda \in \sigma(A) $ and consider a nonzero attack sequence such as $ V(\phi)a(k) = CA^k v $.
	Since $ v \in \mathcal{E}(\lambda) $, i.e., $ v $ is an eigenvector of $ A $, it holds that $ CA^kv = \lambda^k C v $.
	Then, by the relation $ v \in \ker C_\cS $, we have $ C_\mathcal{S} A^k v = \lambda^k C_\cS v  = 0 $.
	For two distinct initial states $ x^1(0) $ and $ x^2(0) \triangleq x^1(0) + v $ and the constructed attack sequence, two output sequences $ y(x^1(0),u,a,\phi,k) $ and $ y(x^2(0),u,0,\phi,k) $ coincide, which implies that $ a(k) $ satisfying $ V(\phi)a(k) = CA^k v $ is undetectable.
	Since now $ V(\phi)a(k) $ has nonzero entries only in the block corresponding to $ \phi_i = \sN $, the security index can be written as
	\begin{align*}
		\alpha_\phi(A, C) &= \min_{a\neq0} \left| \mathrm{blocksupp}_{\phi}(a) \right| \\
		& = \min_{a\neq0} \left| \bigcup_{k \in \mathbb{N}_0} \left\lbrace i \in \mathcal{N} : a_i(k) \neq 0 \right\rbrace \right| \\
		& = \min_{v \in \ker C_\cS \cap \mu(A)} \left| \left\lbrace i \in \mathcal{N}: C_i v \neq 0 \right\rbrace \right|.
	\end{align*}
	Since now $ C_\cS v = 0 $, it can also be presented as 
	\begin{align*}
		\alpha_\phi (A,C) = \min_{v \in \mu(A)}\left| \left\lbrace i \in \mathcal{N} : C_{\{i\}\cup \cS } v \neq 0 \right\rbrace \right|.
	\end{align*}
	
	According to the definition of the security index, the relation $ 3) \Rightarrow 1) $ is trivial, and thus we conclude this proof.
	\endproof
	
	\section{Proof of Corollary \ref{corollary:number_attacks}}
	\label{appendix:proof_corollary:number_attacks}
		First consider $ \alpha_{\phi}(A,C) = + \infty $.
		By definition, $ \alpha_{\phi}(A,C) = + \infty $ means that there is no undetectable attack at all.
		Hence, for any finite $ l $, there is no undetectable attack whose block support size is at most $ l $, and the inequality $ l \geq \alpha_{\phi}(A,C) $ is never satisfied.
		Therefore, both statements are false, so the equivalence holds trivially in this case.
	
		Then, consider $ \alpha_{\phi}(A,C) < + \infty $.
		For necessity, if an undetectable attack $a$ exists under the constraint $| \mathrm{blocksupp}_{\phi}(a) | \leq l$, then, by (\ref{eq:security_index}), we have $ \alpha_\phi(A,C) \leq l $.
		For sufficiency, suppose that $ l \geq \alpha_{\phi}(A,C) $.
		Since $ \alpha_{\phi}(A,C) < + \infty $ means there is an undetectable attack, from (\ref{eq:security_index}), there is an undetectable attack sequence $ a^\star $ such that $ | \mathrm{blocksupp}_{\phi}(a^\star) | = \alpha_{\phi}(A,C) \leq l $.
		Thus, there exists an undetectable attack whose block support size is at most $l$.\endproof
	
	\section{Proof of Lemma \ref{lemma:undetectable}}
	\label{appendix:proof_lemma:undetectable}
	From Proposition~\ref{proposition:undetectable_attacks}, $ \alpha_\phi(A, C) = +\infty $ if and only if $ \ker C_\cS\cap \E(\lambda) = \{0\} $ for all $ \lambda \in \sigma(A) $, which implies that, by the definition of $ E_\lambda $, $ \ker C_\cS\cap \im E_\lambda = \{0\},~\forall \lambda \in\sigma(A) $.
	This is equivalent to $ \ker C_\cS E_\lambda = \{0\},~\forall \lambda \in\sigma(A) $.
	Further, by the full-column rank property of $ E_\lambda $, this is equivalent to $ \rank(C_\cS E_\lambda) = \rank(E_\lambda) = r_\lambda $. \endproof

	\section{Proof of Theorem~\ref{theorem:estimator_performance}}
	\label{appendix:proof_theorem2}
	%
	We first define the following vectors on the states and estimates of all agents:
	\begin{align}
		X(k) & \triangleq \bm{1}_N \otimes x(k) \in \R^{{nN}}, \nonumber \\
		\hat \Xi(k) & \triangleq \left[~\hat \xi_1 (k)^\top,\ldots,\hat \xi_N (k)^\top~\right]^\top \in\R^{{nN}}, \nonumber \\
		\tilde \Xi(k) & \triangleq \left[~\tilde \xi_1 (k)^\top,\ldots,\tilde \xi_N (k)^\top~\right]^\top \in\R^{{nN}}, \nonumber \\
		\bar \Xi(k) & \triangleq \left[~\bar \xi_1 (k)^\top,\ldots,\bar \xi_N (k)^\top~\right]^\top \in\R^{{nN}}.
	\end{align}
	Also, we give
	\begin{align}
		\tilde C & \triangleq \mathrm{blkdiag}\left(C_1,\ldots,C_N\right) \in \R^{p\times {nN}}, \nonumber\\
		S(\phi) & \triangleq \mathrm{blkdiag}\left(S_1(\phi_1),\ldots,S_N(\phi_N)\right) \in \R^{p\times p},\nonumber \\
		S_i(\phi_i)  & \triangleq \left\lbrace \begin{array}{ll}
			0_{p_i}, & \mathrm{if}~\phi_i = \sN, \\
			I_{p_i}, & \mathrm{if}~\phi_i= \sS.
		\end{array}\right. 
	\end{align}
	Note that  $ S(\phi) V(\phi) = 0$ for any $ \phi \in \mathscr{C}^N $ and $  \tilde{C}X(k) = {C}x(k)$.
	We then obtain the following equations regarding the time update and observation update based on (\ref{eq:time_update}) and (\ref{eq:observation_update}):
	\begin{align}
		\bar \Xi(k) &= \left(I_N \otimes {A}\right)\hat \Xi(k-1) + \left(\bm{1}_N \otimes {B} u(k-1)\right) \\
		\tilde \Xi(k) &= \bar \Xi(k) + \tilde{C}^\top S(\phi^\star) \left(y(k) - \tilde{C} \bar \Xi(k)\right).
	\end{align}

	\begin{figure*}[b!]
	\normalsize
	\hrulefill
	\begin{align}
		\label{eq:extended_estimate}
		\hat \Xi(k)& = \left(I_{{nN}} - \omega\left(\mathcal{L}\otimes I_{{n}}\right)\right)^L\tilde \Xi(k) \nonumber \\
		& = \left(I_{{nN}} - \omega\left(\mathcal{L}\otimes I_{{n}}\right)\right)^L \left[\left(I_N \otimes {A}\right)\hat \Xi(k-1) + \left(\bm{1}_N \otimes {B} u(k-1)\right) \right. \nonumber \\
		& \left.\hspace{10mm} + \tilde{C}^\top S(\phi^\star) \left(y(k) - \tilde{C} \left(I_N \otimes {A}\right)\hat \Xi(k-1) - \tilde{C}\left(\bm{1}_N \otimes {B} u(k-1)\right)\right)\right]. \\
		\label{eq:extended_error}
		E(k)  &=\left(I_{{nN}} - \omega\left(\mathcal{L}\otimes I_{{n}}\right)\right)^L \left[\left(I_{{nN}} - \tilde{C}^\top S(\phi^\star) \tilde{C}\right)\left(I_N \otimes {A}\right)E(k-1) \right. \nonumber \\
		& \left.\hspace{10mm} + \left(\tilde{C}^\top S(\phi^\star) \tilde{C} - I_{{nN}}\right)\left(\bm{1}_N \otimes w(k-1)\right) + \tilde{C}^\top S(\phi^\star)v(k)\right].
	\end{align}
	\end{figure*}
	\noindent
	Note that there is no attack vector term $ a(k) $ in this equation since $ S(\phi^\star)V(\phi^\star) = 0 $.
	From (\ref{eq:estimate_consensus}), the extended estimate $ \hat \Xi(k) $ can be obtained in (\ref{eq:extended_estimate}) shown at the bottom of the next page.
	Defining the extended estimation error as $ E(k) \triangleq \hat \Xi(k) - X(k) $, it follows (\ref{eq:extended_error}) shown at the bottom of the next page.
	For the sake of notational simplicity, we define
	\begin{align*}
		M &\!\triangleq\!\left(I_{{nN}}\! -\! \omega\left(\mathcal{L}\!\otimes\! I_{{n}}\right)\right)^L \!\left(I_{{nN}}\!-\! \tilde{C}^\top S(\phi^\star) \tilde{C}\right)\!\left(I_N \!\otimes\! {A}\right), \\
		D_w & \! \triangleq\!\left(I_{{nN}} \!- \!\omega\left(\mathcal{L}\!\otimes \!I_{{n}}\right)\right)^L \!\left(\tilde{C}^\top S(\phi^\star) \tilde{C} \!- \! I_{{nN}}\right),\\
		D_v &\! \triangleq \! \left(I_{{nN}} \!-\! \omega\left(\mathcal{L}\!\otimes\! I_{{n}}\right)\right)^L \tilde{C}^\top S(\phi^\star).
	\end{align*}
	Then, we can simplify the extended error system as
	\begin{align}
		\label{eq:extended_error_2}
		E(k)\!=\! ME(k-1) \!+\! D_w \left(\bm{1}_N \otimes w(k-1)\right) \! + \! D_v v(k).
	\end{align}
	We first seek a sufficient condition such that this extended error system is stable.
	
	For this purpose, we define two orthogonal projection matrices onto the consensus subspace and the orthogonal complement of the consensus subspace, respectively, as follows \cite{2013TSPHachem}:
	\begin{align}
		P_{\sharp} & \triangleq \frac{1}{N}\left(\bm{1}_N \bm{1}_N^\top\right) \otimes I_{{n}} \in \R^{{nN} \times {nN}}, \nonumber \\
		P_{\bot} & \triangleq I_{{nN}} - P_{\sharp} \in \R^{{nN} \times {nN}}.
	\end{align}
	Note that $ \| P_\sharp \| = \|P_\bot \| =  1 $, {$ P_\sharp^2 = P_\sharp $, and $ P_\bot^2 = P_\bot $}.
	Using these matrices, we separate the extended error $ E(k) $ into two variables: $ E_\bot (k) \triangleq P_\bot E(k) $ and $ E_\sharp(k) \triangleq P_\sharp E(k) $, where, by the definition, it holds that $  E_\bot(k) + E_\sharp(k) = E(k) $.
	{Note that, since $ P_\sharp^2 = P_\sharp $ and $ P_\bot^2 = P_\bot $, we obtain $ E_\bot(k)=P_\bot E_\bot(k) $ and $ E_\sharp(k)=P_\sharp E_\sharp(k) $.}
	The separated variables follow the following dynamics:
	\begin{align}
		\label{eq:E_bot_dynamics}
		E_\bot\!(k) &\!= \!P_\bot\! ME(k \! - \! 1) \!+\! \underbrace{P_\bot \! D_w \left(\bm{1}_N \! \otimes \! w(k \! - \! 1)\right)  \! +\! P_\bot \! D_v v(k)}_{\Delta_\bot} \nonumber \\
		&{=\! P_\bot M \left(E_\bot(k-1)+E_\sharp(k-1)\right) + \Delta_\bot} \nonumber \\
		&{=\! \underbrace{P_\bot M P_\bot}_{M_{\bot \bot}} E_\bot(k-1)+\underbrace{P_\bot M P_\sharp}_{M_{\bot \sharp}} E_\sharp(k-1)+ \Delta_\bot} \nonumber \\
		& {= M_{\bot \bot} E_\bot (k-1) + M_{\bot \sharp} E_\sharp(k-1) + \Delta_\bot},
	\end{align}
	\begin{align}
		\label{eq:E_sharp_dynamics}
		E_\sharp(k) &\!=\! P_\sharp ME(k\!-\!1) \!+\! \underbrace{P_\sharp D_w \left(\bm{1}_N \! \otimes \! w(k \! -\!1)\right) \! + \! P_\sharp D_v v(k)}_{\Delta_\sharp}, \nonumber \\
		&{=\! P_\sharp M \left(E_\bot(k-1)+E_\sharp(k-1)\right) + \Delta_\sharp}, \nonumber \\
		&{=\! \underbrace{P_\sharp M P_\bot}_{M_{\sharp \bot}} E_\bot(k-1)+\underbrace{P_\sharp M P_\sharp}_{M_{\sharp \sharp}} E_\sharp(k-1)+ \Delta_\sharp} \nonumber \\
		& {= M_{\sharp \bot} E_\bot (k-1) + M_{\sharp \sharp} E_\sharp(k-1) + \Delta_\sharp},
	\end{align}
	{This implies that their dynamics can be formulated as
	\begin{align}
		\label{eq:E_dynamics}
		\left[\!\!\begin{array}{c}
			E_\bot(k)\\E_\sharp(k)
		\end{array}\!\!\right] = \underbrace{\left[\!\!\begin{array}{cc}
		M_{\bot \bot} & M_{\bot \sharp} \\
		M_{\sharp \bot} & M_{\sharp \sharp}
	\end{array}\!\!\right]}_Z \left[\!\!\begin{array}{c}
		E_\bot(k-1)\\E_\sharp(k-1)
	\end{array}\!\!\right] + \left[\!\!\begin{array}{c}
	\Delta_\bot \\ \Delta_\sharp
\end{array}\!\!\right]\!\!. 
	\end{align}
	Therefore, if $ Z $ is stable, then the dynamics of the extended error $ E(k) $ are also stable.
	A sufficient condition for the stability of $ Z $ is $ \| M_{\bot \bot}\| <1 $, $ \|M_{\sharp \sharp}\| < 1$, and $ \|M_{\bot \sharp}\|\cdot \|M_{\sharp \bot}\| < (1-\|M_{\bot\bot}\|)(1-\|M_{\sharp \sharp}\|)$ (see, e.g., \cite{MatrixAnalysis}).}
	
		
	We first consider {the norm of $ \|M_{\bot \bot}\| $.
	It follows}
	\begin{align*}
		&{\|M_{\bot \bot}\|} \leq  \|P_{\bot} M \| \leq \\ 
		&\| P_{\bot} \!\left(I_{{nN}} \!-\! \omega\left(\mathcal{L} \!\otimes\! I_{{n}}\right)\right)^L \! \|\! \cdot \!\| I_{{nN}}\! -\! \tilde{C}^\top S(\phi^\star) \tilde{C}\| \!\cdot\! \|I_N \!\otimes\! {A}\|
	\end{align*}
	For the terms of $ \tilde C $ and $ {A}$, it can be derived that
	\begin{align*}
		\left\| I_{{nN}}  - \tilde{C}^\top S(\phi^\star) \tilde{C}\right\|  \cdot  \left\|I_N  \otimes  {A}\right\| \leq {\theta_0} \left\| A\right\|,
	\end{align*}
	where $ {\theta_0  \triangleq \max \{1,\max_{i \in \cS^\star}\left\| I_{{n}} - C_i^\top C_i\right\| \}}$.
	Regarding the term of $  \| P_{\bot}\left(I_{{nN}} - \omega\left(\mathcal{L}\otimes I_{{n}}\right)\right)^L\|  $, the eigenvalues of $ I_{{nN}} - \omega\left(\mathcal{L}\otimes I_{{n}}\right) $ are given as $ 1 - \omega \lambda_i^\cL $ with multiplicity $ n $.
	This means that the eigenvalues of $ \left(I_{{nN}} - \omega\left(\mathcal{L}\otimes I_{{n}}\right)\right)^L $ are given as $ \left( 1 - \omega \lambda_i^\cL \right)^L $ with multiplicity $ {n} $.
	Let $q_1\triangleq\bm{1}_N$ be the eigenvector of $\cL$ associated with $\lambda_1^\mathcal L=0$.
	For any $z \in \R^n$, $P_\perp(q_1 \otimes z) = 0$.
	Thus, it follows that
	\begin{align*}
		P_{\bot}\left(I_{{nN}} - \omega\left(\mathcal{L}\otimes I_{{n}}\right)\right)^L\left(q_1 \otimes z \right) = 0,
	\end{align*}
	which indicates that the eigenvalue of $ \lambda_1^\cL= 0 $ is negligible. 
	As a result, we obtain
	\begin{align}
		\label{eq:P_bot_01}
		\left\| P_{\bot}\left(I_{{nN}} - \omega\left(\mathcal{L}\otimes I_{{n}}\right)\right)^L\right\| = \max_{i=2,\ldots,N}\left| 1 - \omega \lambda_i^\cL\right|^L.
	\end{align}
	Let $ \omega = 2/(\lambda_{2}^\cL + \lambda_{\max}^\cL) $, then we have
	\begin{align}
		\label{eq:P_bot_02}
		\max_{i=2,\ldots,N}\left| 1 - \omega \lambda_i^\cL\right|^L = \left( \dfrac{\lambda_{\max}^\cL- \lambda_2^\cL}{\lambda_{\max}^\cL+ \lambda_2^\cL}\right)^L {= \gamma_\bot^L},
	\end{align}
	where $ \gamma_{\bot} \triangleq \left(\lambda_{\max}^\cL - \lambda_{2}^\cL\right)/\left(\lambda_{\max}^\cL + \lambda_{2}^\cL\right) $.
	{Consequently, $ \| M_{\bot \bot}\| \leq \gamma_\bot^L \theta_0 \| A\| $, which implies that, if $ \gamma^L_\bot \theta_0 \|A\| < 1 $, then $ \|M_{\bot \bot}\| < 1 $.}
	
	{For the norm of $ \|M_{\bot \sharp}\| $, it follows that $ \|M_{\bot \sharp}\| \leq \|P_\bot M\| $.
	Therefore, from the above results, we obtain $ \| M_{\bot \sharp}\| \leq \gamma_\bot^L \theta_0 \| A\| $.}

	{For the norm of $ \|M_{\sharp \bot}\| $, it holds that $ \|M_{\sharp \bot}\| \leq \|P_\sharp M\| $.}
	From the relation that $  \bm{1}_N^\top \mathcal{L} = 0 $, we have $ P_\sharp \left(\mathcal{L}\otimes I_{{n}}\right) = 0$, which indicates $ P_\sharp\left(I_{{nN}} - \omega\left(\mathcal{L}\otimes I_{{n}}\right)\right)^L = P_\sharp $.
	Hence, we have
	\begin{align*}
		{\| M_{\sharp \bot}\| \!\leq} \|P_{\sharp}M\| \!\leq\! \left\| I_{{nN}}  \!-\! \tilde{C}^\top S(\phi^\star) \tilde{C}\right\| \! \cdot \! \left\|I_N \! \otimes \! {A}\right\| \!\leq\! {\theta_0} \!\left\| A\right\|\!.
	\end{align*}
	
	{Finally, we consider the norm of $ \|M_{\sharp \sharp}\| $.
   	By construction, $ E_{\sharp}(k) $ can be written as 
    	\begin{align}
    		\label{eq:e_c(k)}
    		E_\sharp(k) & = P_\sharp E(k) = \left(\frac{1}{N}\left(\bm{1}_N \bm{1}_N^\top\right) \otimes I_{{n}}\right)E(k) \nonumber\\
    		& = \bm{1}_N \otimes \underbrace{\left(\frac{1}{N}\sum_{i=1}^{N} \left(\hat \xi_i(k) - x(k)\right)\right)}_{e_c(k)},
    	\end{align}
    which implies $ E_\sharp(k) $ can be represented using $ e_c(k) \in \R^n $.
	Using this relation, (\ref{eq:E_sharp_dynamics}) can be reformulated as
	\begin{align}
		\label{eq:theorem_e_c}
		 E_\sharp(k) 
		\!=\!  M_{\sharp \sharp} \left(\bm{1}_N \!\otimes \!e_c(k-1)\right)\!+\!M_{\sharp \bot}E_\bot(k-1)\!+\!\Delta_\sharp.
	\end{align}
	Thus, for the term $ M_{\sharp \sharp} \left(\bm{1}_N \!\otimes \!e_c(k-1)\right) $ of (\ref{eq:theorem_e_c}), we have
	\begin{align}
	\label{eq:theorem_P_sharp_M}
	&M_{\sharp \sharp} \left(\bm{1}_N \!\otimes \!e_c(k-1)\right) \nonumber\\
	 &= P_\sharp\left(I_{{nN}}\!-\! \tilde{C}^\top S(\phi^\star) \tilde{C}\right)\!\left(I_N \!\otimes\! {A}\right)\!P_\sharp\!\left(\bm{1}_N \!\otimes \!e_c(k-1)\right) \nonumber\\
	 & = P_\sharp\left(I_{{nN}}\!-\! \tilde{C}^\top S(\phi^\star) \tilde{C}\right)\left(\bm{1}_N \!\otimes \!(Ae_c(k-1))\right) \nonumber \\
	 & = P_\sharp \left[\begin{array}{c}
	 G_1A \\ \vdots \\ G_NA
	 \end{array}\right]e_c(k-1),
	\end{align}
	where
	\begin{align*}
		G_i \triangleq \left\lbrace \begin{array}{ll}
			I_n - C_i^\top C_i, & i \in \cS^\star, \\
			I_n, & i \notin \cS^\star.
		\end{array}\right.
	\end{align*}
	By the definition of $ P_\sharp $, (\ref{eq:theorem_P_sharp_M}) can be written as
	\begin{align}
		&M_{\sharp \sharp} \left(\bm{1}_N \!\otimes \!e_c(k-1)\right)\nonumber \\
		&= \bm{1}_N \otimes \left(\left(I_n - \dfrac{1}{N}\sum_{i \in \mathcal{S}^\star }C_i^\top C_i\right)A\cdot e_c(k-1)\right) \nonumber \\ 
		& = \bm{1}_N \otimes \left( F e_c(k-1) \right),
	\end{align}
	where $ F \triangleq (I_n - \frac{1}{N}\sum_{i \in \mathcal{S}^\star }C_i^\top C_i)A $.
	Therefore, (\ref{eq:theorem_e_c}) can be reformulated as
	\begin{align}
	\label{eq:E_sharp}
	E_\sharp(k)  & =\bm{1}_N \!\otimes\! \left( F e_c(k-1) \right) +M_{\sharp \bot}E_\bot(k-1)+\Delta_\sharp \nonumber\\
	& = \left(J \!\otimes \!F \right)\left(\bm{1}_N \!\otimes\! e_c(k-1)\right)+M_{\sharp \bot}E_\bot(k-1)+\Delta_\sharp \nonumber\\
	& = \left(J \!\otimes \!F \right)E_\sharp(k-1)+M_{\sharp \bot}E_\bot(k-1)+\Delta_\sharp,
	\end{align}
	where $ J \triangleq \frac{1}{N} \bm{1}_N \bm{1}_N^\top \in \R^{N \times N} $.
	This implies that $ M_{\sharp \sharp} = J \otimes F $ and, if $ \|J \otimes F \| = \| F \| < 1 $, then $ \|M_{\sharp \sharp}\| < 1 $.}
	
	{
	Combining the above results, if $ \gamma_\bot^L \theta_0 \|A\| < 1 $, $ \|F\|<1 $, and
	\begin{align*}
		\gamma_{\bot}^L \theta_0^2 \|A\|^2 < \left(1-\gamma_\bot^L \theta_0 \|A\|\right)\left(1-\|F\|\right),
	\end{align*}
	then the block matrix $ Z $ is stable, which implies the dynamics of the extended error $ E(k) $ are also stable.
	The condition $ \gamma_\bot^L \theta_0 \|A\| < 1 $ can be reformulated as (\ref{eq:L}).
	Therefore, if (\ref{eq:L})--(\ref{eq:last_condition}) hold, then the dynamics of $ E(k) $ are stable.}

	We next consider the asymptotic upper bound of $ E(k) $.
	{From (\ref{eq:E_dynamics}) and the above results, we obtain
	\begin{align}
		\underbrace{\left[\!\!\!\begin{array}{c}
			\|E_\bot\!(k\!+\!1)\| \\ \| E_\sharp(k\!+\!1)\|
		\end{array}\!\!\!\right]}_{\psi(k+1)}\! \leq \!\underbrace{\left[\!\!\!\begin{array}{cc}
		\gamma_\bot^L \theta_0 \|A\| \!& \!\gamma_\bot^L \theta_0 \|A\|\\
		\theta_0 \|A\| \!& \!\|F\|
	\end{array}\!\!\!\right]}_{\bar Z}\!\!	\underbrace{\left[\!\!\!\begin{array}{c}
	\|E_\bot\!(k)\| \\ \| E_\sharp(k)\|
\end{array}\!\!\!\right]}_{\psi(k)}\!\! +\!\! 	\underbrace{\left[\!\!\!\begin{array}{c}
\|\Delta_\bot\| \\ \| \Delta_\sharp \|
\end{array}\!\!\!\right]}_{\Delta_z}\!,
\end{align}
where $ \psi(k) \in \R^{2} $, $ \bar Z \in \R^{2 \times 2} $, and $ \Delta_z \in \R^2 $.
Therefore, we have
\begin{align*}
	\psi(k) \leq \bar Z^k \psi(0) + \sum_{j=0}^{k-1} \bar Z^j \Delta_z.
\end{align*}
If (\ref{eq:L})--(\ref{eq:last_condition}) hold, the $ \bar Z $ is Schur stable, and hence $ \sum_{j=0}^{\infty}\bar Z^j = (I-\bar Z)^{-1} $.
Consequently, we obtain 
\begin{align}
	\label{eq:psi_limit}
	\limsup_{k \rightarrow \infty} \psi(k) \leq \sum_{j=0}^{\infty} \bar Z^j \Delta_z = \left(I-\bar Z\right)^{-1} \Delta_z.
\end{align}
To give the upper bound of $ E(k) $, thus, this proof tackles providing $ \left(I-\bar Z\right)^{-1} $ and the upper bound of $ \Delta_z $.}

{First, for the upper bound of $ \Delta_z $, consider $ \| \Delta_\bot \| $.}
We obtain
	\begin{align*}
		&\left\|  P_\bot D_w \left(\bm{1}_N \otimes w(k-1)\right) \right\| \\
		& \leq 	\left\|  P_\bot D_w \right\| \cdot \left\|\left(\bm{1}_N \otimes w(k-1)\right)  \right\| \nonumber \\
		& \leq \left\|  P_{\bot}\!\left(I_{{nN}} \!-\! \omega\left(\mathcal{L}\otimes I_{{n}}\right)\right)^L  \right\| \nonumber \\
		& \hspace{15mm} \! \cdot \! \left\| \tilde{C}^\top S(\phi^\star) \tilde{C} - I_{{nN}} \right\| \!\cdot \!\left({\sqrt{N} \delta_w}\right)  \leq {\gamma_{\bot}^L \theta_0 \sqrt{N} \delta_w}.
	\end{align*}
	Also, we have
	\begin{align*}
		\left\|P_\bot D_v v(k)  \right\| & \leq \left\|P_\bot D_v   \right\| \cdot \left\| v(k)  \right\| \nonumber \\
		& \leq  \gamma_\bot^L \cdot \left\| \tilde C^\top S(\phi^\star) \right\|\cdot \delta_v\leq {\gamma_\bot^L \nu_0 \delta_v},
	\end{align*}
	where $ \nu_0 \triangleq \max_{i \in \cS^\star}\left\| C_i\right\| $.
	Therefore, it follows that
	\begin{align}
		\label{eq:Delta_bot}
		\left\| \Delta_\bot \right\|\leq \gamma_{\bot}^L ( \underbrace{{\theta_0 \sqrt{N}\delta_w} + \nu_0{\delta_v}}_D)~{= \gamma_\bot^L D}.
	\end{align}

	For $ \|\Delta_\sharp\|  $, we have
	\begin{align*}
		&\left\|  P_\sharp D_w \left(\bm{1}_N \otimes w(k-1)\right)  \right\| \nonumber \\
		& \leq 	\left\|  P_\sharp D_w \right\| \cdot \left\|\left(\bm{1}_N \otimes w(k-1)\right)  \right\| \nonumber \\
		& \overset{(a)}{\leq} \left\| P_\sharp \left(\tilde{C}^\top S(\phi^\star) \tilde{C} - I_{{nN}}\right) \right\| \cdot \left({\sqrt{N} \delta_w}\right) \leq {\theta_0 \sqrt{N} \delta_w},
	\end{align*}
	where inequality (a) stems from the relation of $ P_\sharp\left(I_{{nN}} - \omega\left(\mathcal{L}\otimes I_{{n}}\right)\right)^L = P_\sharp $.
	Similarly, it follows that
	\begin{align*}
		\left\|P_\sharp D_v v(k)  \right\| & \leq \left\|P_\sharp D_v   \right\| \cdot \left\| v(k)  \right\| \leq \left\| P_\sharp \left(\tilde{C}^\top S(\phi^\star) \right)\right\| \cdot {\delta_v}\nonumber \\
		& \leq  \left\| \tilde{C}^\top S(\phi^\star) \right\| \cdot {\delta_v}  = \nu_0{\delta_v},
	\end{align*}
	which indicates
	\begin{align}
		\label{eq:Delta_sharp}
		\left\| \Delta_\sharp \right\|\leq {\theta_0 \sqrt{N} \delta_w} + \nu_0 {\delta_v}~{=D}.
	\end{align}
	{Therefore, $ \Delta_z $ follows
	\begin{align}
		\Delta_z = \left[\begin{array}{c}
			\|\Delta_\bot\| \\ \| \Delta_\sharp \|
		\end{array}\right] \leq \left[\begin{array}{c}
		\gamma_\bot^L D\\ D
	\end{array}\right].
\end{align}}

	{For (\ref{eq:psi_limit}), next considering $ \left(I-\bar Z\right)^{-1} $, it holds that
	\begin{align}
		\left(I-\bar Z\right)^{-1} = \dfrac{1}{\zeta}\left[\begin{array}{cc}
			1-\|F\| & \gamma_\bot^L \theta_0 \|A\| \\
			\theta_0 \|A\| & 1 - \gamma_\bot^L \theta_0 \|A\|
		\end{array}\right],
	\end{align}
	where $ \zeta \triangleq \left(1-\gamma_\bot^L \theta_0 \|A\|\right)\left(1-\|F\|\right)-\gamma_{\bot}^L \theta_0^2 \|A\|^2 $.
	Therefore, we obtain
	\begin{align}
		\limsup_{k \rightarrow \infty} \psi(k)\leq \dfrac{1}{\zeta}\left[\!\!\!\begin{array}{cc}
			1-\|F\| & \gamma_\bot^L \theta_0 \|A\| \\
			\theta_0 \|A\| & 1 - \gamma_\bot^L \theta_0 \|A\|
		\end{array}\!\!\!\right]\left[\!\!\!\begin{array}{c}
		\gamma_\bot^L D\\ D
	\end{array}\!\!\!\right].
	\end{align}
	Finally, for the asymptotic upper bound of $ E(k) $, we have
	\begin{align}
		\limsup_{k \rightarrow \infty}\|E(k)\| & \leq \limsup_{k \rightarrow \infty} \left(\|E_\bot(k)\|+\|E_\sharp(k)\|\right) \nonumber\\
		& = \limsup_{k \rightarrow \infty} \bm{1}_2^\top \psi(k) \nonumber \\
		& \leq \bm{1}_2^\top \!\!\cdot\! \left(\!\dfrac{1}{\zeta}\!\left[\!\!\!\begin{array}{cc}
			1-\|F\| & \gamma_\bot^L \theta_0 \|A\| \\
			\theta_0 \|A\| & 1 - \gamma_\bot^L \theta_0 \|A\|
		\end{array}\!\!\!\right]\!\!\left[\!\!\!\begin{array}{c}
			\gamma_\bot^L D\\ D
		\end{array}\!\!\!\right]\!\right) \nonumber \\
	&= \dfrac{D\left(1+\gamma_\bot^L\left(1-\|F\|+\theta_0 \|A\|\right)\right)}{\zeta}.
	\end{align}}
	This relation holds for all $ e_i(k) ,~i \in \mathcal{V} $, which concludes the proof.\endproof
%
	
	\section{Proof of Proposition \ref{proposition:control}}
	\label{appendix:proof_proposition1}
	For the dynamics of the control error $ \tilde e_i(k) $, we obtain
	\begin{align}
		\label{eq:proposition_3_1}
		\tilde e_i(k\!+\!1) & \!=\! x_i(k\!+\!1) \!-\! x^*_i(k\!+\!1) \nonumber \\
		& \!=\! {A_i}x_i(k) \!+\! {B_i}u_i(k) \!+\! w_i(k) \!-\! \left({A_ix^*_i(k) \!+\! B_i u^*_i(k)}\right) \nonumber \\
		&\! =\! \left({A_i \!-\! B_iK_i}\right)\tilde e_i(k) \!-\! {B_iK_i} e_i(k) \!+\! w_i(k).
	\end{align}
	{Since $ {\rho(A_i - B_i K_i)}< 1 $, (\ref{eq:proposition_3_1}) is input-to-state stable with respect to $ e_i $ and $ w_i $. Hence, using the standard bound for stable linear systems with bounded inputs, we obtain}
	\begin{align}
		\limsup_{k \rightarrow \infty} \left\| \tilde e_i (k) \right\| \leq &{\left(\sum_{t=0}^{\infty} \left\| (A_i - B_i K_i)^t \right\|\right) }\nonumber\\
		&~~~~~~{\cdot \left(\left\| B_iK_i\right\|\Delta_e^\infty + \delta^w_i\right)},
	\end{align}
	where $ \Delta_e^\infty  $ is the asymptotic upper bound of the estimation error.
	By Theorem~\ref{theorem:estimator_performance}, if {(\ref{eq:L})--(\ref{eq:last_condition}) hold} with $ \omega =  2/\left(\lambda_{2}^\cL + \lambda_{\max}^\cL \right)$, then the upper bound is given by (\ref{eq:error}) {for all $ i \in \V $}, which concludes the proof. \endproof

	\begin{IEEEbiographynophoto}
		{Takumi Shinohara} (Member, IEEE) received the B.E., M.E., and Ph.D. degrees from Keio University, Tokyo, Japan, in 2016, 2018, and 2024, respectively. From 2018 to 2025, he was a consultant at Mitsubishi Research Institute, and from 2024 to 2025, he was a visiting researcher at Keio University. Since 2025, he has been a Postdoctoral Researcher with the Department of Decision and Control Systems at KTH Royal Institute of Technology, Stockholm, Sweden.
		His research interests include security and resilience of networked control systems.
	\end{IEEEbiographynophoto}

	\begin{IEEEbiographynophoto}
		{Karl Henrik Johansson} (Fellow, IEEE) received the M.Sc. degree in electrical engineering and the Ph.D. degree in automatic control from Lund University, Lund, Sweden, in 1992 and 1997, respectively.
		
		He is a Swedish Research Council Distinguished Professor in electrical engineering and computer science with the KTH Royal Institute of Technology, Stockholm, Sweden, and the Founding Director of Digital Futures. He has held Visiting Positions with UC Berkeley, Caltech, NTU, and other prestigious institutions. His research interests include networked control systems and cyber-physical systems with applications in transportation, energy, and automation networks.
		
		Dr. Johansson was the recipient of numerous best paper awards and various distinctions from IEEE, IFAC, and other organizations, for his scientific contributions, and also Distinguished Professor by the Swedish Research Council, Wallenberg Scholar by the Knut and Alice Wallenberg Foundation, Future Research Leader by the Swedish Foundation for Strategic Research, triennial IFAC Young Author Prize and IEEE CSS Distinguished Lecturer, and IEEE CSS Hendrik W. Bode Lecture Prize. His extensive service to the academic community includes serving as President of the European Control Association, IEEE CSS Vice President Diversity, Outreach \& Development, and Member of IEEE CSS Board of Governors and IFAC Council. He was on the editorial boards of Automatica, IEEE Transactions on Automatic Control, IEEE Transactions on Control of Network Systems, and many other journals. He has also been a Member of the Swedish Scientific Council for Natural Sciences and Engineering Sciences. He is a Fellow of the Royal Swedish Academy of Engineering Sciences.
	\end{IEEEbiographynophoto}

	\begin{IEEEbiographynophoto}
		{Henrik Sandberg} (Fellow, IEEE) is Professor and Deputy Head at the Department of Decision and Control Systems, KTH Royal Institute of Technology, Stockholm, Sweden. He received the M.Sc. degree in engineering physics and the Ph.D. degree in automatic control from Lund University, Lund, Sweden, in 1999 and 2004, respectively. From 2005 to 2007, he was a Postdoctoral Scholar at the California Institute of Technology, Pasadena, USA. In 2013, he was a Visiting Scholar at the Laboratory for Information and Decision Systems (LIDS) at MIT, Cambridge, USA. He has also held visiting appointments at the Australian National University and the University of Melbourne, Australia. His current research interests include security of cyber-physical systems, power systems, model reduction, and fundamental limitations in control. Dr. Sandberg was a recipient of the Best Student Paper Award from the IEEE Conference on Decision and Control in 2004, an Ingvar Carlsson Award from the Swedish Foundation for Strategic Research in 2007, and a Consolidator Grant from the Swedish Research Council in 2016. He has served on the editorial boards of IEEE Transactions on Automatic Control and the IFAC Journal Automatica. He is a Member of the IEEE CSS Board of Governors, is a Fellow of the Royal Swedish Academy of Engineering Sciences, and a Fellow of the IEEE.
	\end{IEEEbiographynophoto}
	
\end{document}